\documentclass[11pt,a4paper,preprintnumbers]{article}

\usepackage{jheppub}
\usepackage{graphicx}
\usepackage{amsmath}
\usepackage{xspace}
\usepackage{xcolor}
\usepackage{float}
\usepackage{comment}
\usepackage{subcaption}
\usepackage{textcomp}
\usepackage{placeins}
\usepackage{bm}
\usepackage{pstricks}
\usepackage{color}
\usepackage{braket}
\usepackage{slashed}

\newcommand{\eq}[1]{eq.~\eqref{eq:#1}}
\newcommand{\eqs}[2]{eqs.~\eqref{eq:#1} and \eqref{eq:#2}}
\renewcommand{\sec}[1]{sec.~\ref{sec:#1}}

\newcommand{\fig}[1]{fig.~\ref{fig:#1}}

\newcommand{\app}[1]{App.~\ref{app:#1}}

\newcommand{\tab}[1]{Tab.~\ref{tab:#1}}

\newcommand{\ord}[1]{\mathcal{O}(#1)}

\newcommand{\rescaletwoplots}{0.49\textwidth}
\newcommand{\hspacebetweentwoplots}{\hspace*{.02\textwidth}}

\newcommand{\rescalethreeplots}{0.32\textwidth}

\newcommand{\spaceabovefigurecaption}{\vspace*{-2ex}}
\newcommand{\spacebelowfigurecaption}{\vspace*{0ex}}
\newcommand{\as}{\alpha_{\rm s}}

\newcommand{\betat}{\beta_t}

\newcommand{\Tau}{\mathcal{T}}

\newcommand{\GeV}{\,\mathrm{GeV}}

\newcommand{\nn}{\nonumber}

\newcommand{\NS}{\mathrm{NS}}

\newcommand{\cusp}{\mathrm{cusp}}

\newcommand{\ttbar}{$t\bar{t}$~}
\newcommand{\ttbarj}{$t\bar{t}$+jet~}

\newcommand{\tausq}{l^+}
\newcommand{\dvol}{\Omega_{d-4}}

\newcommand{\de}{\mathrm{d}}

\newcommand{\geneva}{\textsc{Geneva}\xspace}
\newcommand{\minnlops}{\textsc{MiNNLO}$_{\textsc{PS}}$\xspace}

\newcommand{\mcfm}{\textsc{Mcfm}\xspace}

\newcommand{\yttbar}{y_{t\bar{t}}}

\renewcommand{\thesection}{\arabic{section}}

\allowdisplaybreaks[2]

\newcommand{\xam}{x_a^-}
\newcommand{\xbm}{x_b^-}

\begin{document}


\title{Zero-jettiness resummation for top-quark pair production at the LHC}
\preprint{DESY 21-186}

\author[a]{Simone Alioli,}

\author[a]{Alessandro Broggio}

\author[a,b]{and Matthew A.~Lim}

\affiliation[a]{Universit\`{a} degli Studi di Milano-Bicocca \& INFN, Piazza della Scienza 3, Milano 20126, Italy\vspace{0.5ex}}
\affiliation[b]{Deutsches Elektronen-Synchrotron DESY, Notkestr. 85, 22607 Hamburg, Germany\vspace{0.5ex}}
\emailAdd{simone.alioli@unimib.it}
\emailAdd{alessandro.broggio@unimib.it}
\emailAdd{matthew.lim@desy.de}

\date{\today}

\abstract{ We study the resummation of the 0-jettiness resolution variable
  $\Tau_0$ for the \mbox{top-quark} pair production process in hadronic
  collisions.  Starting from an effective theory framework we derive a
  factorisation formula for this observable which allows its resummation at any
  logarithmic order in the $\Tau_0\to 0$ limit.  We then calculate the
  $\mathcal{O}(\alpha_s)$ corrections to the soft function matrices and, by
  employing renormalisation group equation methods, we obtain the ingredients
  for the resummation formula up to next-to-next-to-leading logarithmic
  (NNLL) accuracy. We study the impact of these corrections to the
  0-jettiness distribution by comparing predictions at different accuracy orders: NLL,
  NLL$^\prime$, NNLL and approximate NNLL$^\prime$ (NNLL$^\prime_{\text{a}}$).
  We match these results to the corresponding fixed order calculations both at
  leading order and next-to-leading order for the \ttbarj production
  process, obtaining the most accurate prediction of the 0-jettiness
  distribution for the top-quark pair production process at
  NNLL$^\prime_{\text{a}}$+NLO accuracy.  }

\maketitle
\flushbottom

\section{Introduction}
\label{sec:intro}

Top-quark pair production is a process of enormous interest at the LHC and at future hadron colliders. The top-quark is the most massive of the Standard Model (SM) particles and couples strongly to the Higgs boson -- studies of its properties are therefore vital in order to test the SM at the precision frontier and probe the nature of the observed Higgs sector. The top sector is also of interest from the perspective of constraining higher dimensional operators in the Standard Model Effective Theory, as well as for fitting parton distribution functions. For all of these purposes, it is necessary to have predictions for the main top-quark production channel, i.e. pair production, at the highest possible accuracy.

There is an abundance of experimental measurements of the \ttbar process at all LHC run energies and with a range of luminosities, for example Refs.~\cite{CMS:2016hbk,CMS:2017xrt,CMS:2017XIO,CMS:2018adi,CMS:2018tdx,CMS:2020tvq,CMS:2021vhb,ATLAS:2019hxz,ATLAS:2019hau,ATLAS:2020aln,ATLAS:2020ccu}. Such measurements are important for extracting properties of the SM such as the strong coupling $\alpha_s$ and the top-quark mass~\cite{Klijnsma:2017eqp,Cooper-Sarkar:2020twv}.

The first predictions for the \ttbar total cross section at next-to-next-to-leading order (NNLO) in quantum chromodynamics (QCD) appeared in Refs.~\cite{Barnreuther:2012wtj,Czakon:2012pz,Czakon:2013goa}. These were followed by predictions for differential distributions of stable top-quarks computed in the \textsc{Stripper} local subtraction framework~\cite{Czakon:2015owf,Czakon:2016dgf,Czakon:2016ckf,Czakon:2017dip}. The two-loop amplitudes computed for the aforementioned calculations were later reused in a new calculation at NNLO via the $q_T$-slicing approach~\cite{Catani:2019iny,Catani:2019hip,Catani:2020tko}. NNLO corrections to both top-quark production and decay were combined in the narrow-width approximation in Refs.~\cite{Behring:2019iiv,Czakon:2020qbd}, while NLO electroweak (EW) corrections (first computed in Refs.~\cite{Bernreuther:2005is,Kuhn:2006vh,Denner:2016jyo}) were combined with NNLO QCD corrections for stable tops in Ref.~\cite{Czakon:2017wor}.

In addition to calculations at fixed order (FO) in perturbation theory, resummed
calculations have also appeared for a variety of observables for this
process. Following the computation of the relevant anomalous dimensions in
Refs.~\cite{Ferroglia:2009ep,Ferroglia:2009ii}, threshold resummation at
next-to-next-to-leading logarithmic accuracy (NNLL) was achieved via
Soft-Collinear Effective Theory (SCET). This was carried out in different
kinematic regimes -- the pair invariant mass kinematics were studied in
Ref.~\cite{Ahrens:2010zv}, while the treatment of the 1-particle inclusive
kinematics appeared in Ref.~\cite{Ahrens:2011mw}. Within this paradigm, more
differential results including approximate NNLO corrections to the production
and NLO (NNLO) corrections to the decay appeared in
Refs.~\cite{Broggio:2014yca,Gao:2017goi}, while studies of boosted top-quark
production matched to both NLO and NNLO calculations were carried out in
Refs.~\cite{Pecjak:2016nee,Czakon:2018nun}. The transverse momentum of the
\ttbar pair was resummed within SCET in Ref.~\cite{Li:2013mia}, while the
framework for performing the same resummation in direct QCD was later
established in Ref.~\cite{Catani:2018mei}.

Recently, NNLO predictions were matched to a parton shower in
Ref.~\cite{Mazzitelli:2020jio} using the \minnlops
method~\cite{Monni:2019whf}, allowing the generation of events at NNLO+PS
accuracy. This work utilised the factorisation theorem of
Ref.~\cite{Catani:2018mei} and was made possible by the calculation of the
necessary perturbative ingredients in Ref.~\cite{Catani:2019iny} as well as
the POWHEG implementations of the \ttbar and \ttbarj
processes~\cite{Frixione:2007nw,Alioli:2011as}.

In this work, we develop a resummation framework using the 0-jettiness
variable $\Tau_0$ for the production of a \ttbar pair at the LHC, which is a
required ingredient for the creation of a NNLO+PS event generator in the
\geneva formalism~\cite{Alioli:2015toa} when using $\Tau_0$ as resolution
variable.\footnote{Recently the \geneva framework has also been implemented
  using the transverse momentum of the colour singlet as resolution variable
  for the DY production process~\cite{Alioli:2021qbf}. Given that the
  factorisation theorem for the transverse momentum of the $t \bar{t}$ pair
  has already been established in Ref.~\cite{Li:2013mia} and the relevant soft
  functions have been calculated at two-loop order in
  Ref.~\cite{Angeles-Martinez:2018mqh}, one can envision a similar development
  also for $t \bar{t}$ production.  We leave this extension to future
  work. }

The $N$-jettiness variable $\Tau_N$ was first introduced in Ref.~\cite{Stewart:2010tn}
as a means by which to discriminate between resolved emissions with different
jet multiplicities. Given an $M$-particle phase space point $\Phi_M$ with $M
\ge N$, it is defined as
\begin{align}
\label{eq:TauN}
\Tau_N(\Phi_M) = \sum_k \textrm{min}
\big\{\hat{q}_a\cdot p_k, \hat{q}_b\cdot p_k, \hat{q}_1\cdot
p_k,\ldots, \hat{q}_N\cdot p_k \big\} \, ,
\end{align}
where \mbox{$\hat{q}_i=n_i=(1,\vec{n}_i)$} are light-like reference
vectors parallel to the beam and jet directions, defined in the rest
frame of the Born-level final state, and the index $k$ runs over all
strongly-interacting final-state particles. The limit $\Tau_N\to 0$
describes an event with $N$ pencil-like hard jets, where the
unresolved emissions can either be soft or collinear to the
final-state jets or to the beams.\footnote{Different procedures can be devised
to identify the jet directions $\hat{q}_i $ that truly minimise the
sum -- for example, one can identify the jet directions with the axis
found by a jet clustering algorithm. The differences between
these alternative definitions are usually of nonsingular nature.
Given that the only directions which are important for $\Tau_0$ are those of
the beams and there is no ambiguity in their definition, we
will not be concerned with this issue for the rest of this paper.
}

A factorisation theorem for the
$0$-jettiness, valid for colour-singlet production, was derived in
Ref.~\cite{Stewart:2009yx} via SCET methods. This allowed the
resummation of logarithms of $\Tau_0/Q$ for some hard scale $Q$, and
has been the basis for the majority of practical implementations of
the \geneva framework thus far. In the \geneva approach, a resummed calculation in a suitable resolution
variable is combined with the appropriate fixed order predictions and matched
to a parton shower, thus providing accurate predictions across the whole 
phase space. The method has been successfully applied to several processes
involving colour-singlet particles in the final state (or coloured particles
produced from decaying colour singlets) up to N$^3$LL + NNLO
accuracy~\cite{Alioli:2019qzz,Alioli:2020fzf,Alioli:2020qrd,Alioli:2021qbf,Alioli:2021egp,Cridge:2021hfr}. This
work paves the way for the first \geneva implementation for the production of
massive coloured particles in hadronic collisions.

The $N$-jettiness can also be used as a slicing parameter for fixed order calculations~\cite{Gaunt:2015pea}.
In particular, a $0$-jettiness subtraction has been used to obtain NNLO predictions for an array of
colour-singlet production processes in \mcfm~\cite{Boughezal:2016wmq,Campbell:2019dru}. The method has also recently
been extended to N$^3$LO in Ref.~\cite{Billis:2019vxg}.  The results presented in this
work provide an avenue towards a future application of the $0$-jettiness
subtraction to the production of heavy quark pairs at NNLO accuracy.

In addition to its intrinsic utility for the \geneva or \mcfm implementations,
0-jettiness  is an important observable to disentangle the effects of
underlying-event and multi-parton interactions in hadronic collisions.
The reason is that contributions from particles in the forward and backward
directions are suppressed with respect to particles emitted at central
pseudorapidities. Therefore, measuring 0-jettiness, or beam-thrust~\cite{ATLAS:2016hjr}, and comparing it with other observables like the scalar sum of all transverse
momenta gives a handle on the hadronic activity from initial-state radiation.

When using the $N$-jettiness as a resolution variable for processes involving massive coloured
particles, there are two possible choices one could make. The first option is
to extend the definition in \eq{TauN} to include the velocities of the
top-quarks as additional reference directions, i.e. to supplement the
light-like $\hat{q}_i$ with two additional massive vectors. This is similar to the approach
taken in Refs.~\cite{Fleming:2007qr,Fleming:2007xt,Bachu:2020nqn}, which also resums logarithms arising from 
radiation collinear to the heavy quarks.
This is the appropriate treatment when the top-quarks are in the boosted regime, with the top-quark pair invariant mass
 $M\gg m_t$. 

Alternatively, one
could exclude the final-state top-quarks from the sum in
\eq{TauN}. In this way, additional radiation is never clustered together with
a top-quark, which ultimately is not a stable particle. In the soft and
collinear limit and provided that the top-quarks are not highly boosted, the two choices are equivalent, in the sense that all
singularities are captured as $\Tau_N\to 0$ in both cases.\footnote{In this regime, emissions collinear to the top-quark are not divergent. The large quark mass acts as a cut-off, preventing any potential enhancement.} In this work we make the second of the two choices, and resum the 0-jettiness so-defined for the \ttbar process. 

The paper is organised as follows. In \sec{FacTau}, we discuss the
factorisation theorem in SCET and the calculation of the necessary
perturbative ingredients. We then examine the resummation of logarithms via
renormalisation group (RG) evolution in \sec{resummation} and provide expressions
for the resummed $\Tau_0$ distribution at various resummation orders. We
present numerical results for the resummed $\Tau_0$ distribution in
\sec{resumresults}, and finally report our conclusions in \sec{conc}.

\section{Zero-jettiness factorisation for top-quark pair production}
\label{sec:FacTau}
In this section we discuss the factorisation theorem for $\Tau_0$ and the
calculation of the perturbative ingredients necessary to achieve NNLL and
approximate NNLL$^\prime$ accuracy. We present the factorisation theorem for
the \ttbar process, whose complete derivation is detailed in
\app{Hvqfactheorem}. We also report on the calculation of the relevant soft
function at one-loop order, which is a necessary ingredient beginning at
NLL$^\prime$ accuracy.

\subsection{Process definition}
We consider the creation of a $t \bar{t}$ pair from the collision of two
protons in the process
\begin{align}
  P_1(P_a) + P_2(P_b)\to t(p_3)+\bar{t}(p_4)+X(p_X)
\end{align}
which receives contributions from the quark and gluon-initiated Born level partonic channels
\begin{align}
  q(p_1)+\bar{q}(p_2) &\to t(p_3)+\bar{t}(p_4) \quad + \quad (q \leftrightarrow \bar{q}) \,,\nn\\
  g(p_1)+g(p_2) &\to t(p_3)+\bar{t}(p_4)\,,
\end{align}
where $p_1 = x_1 P_a\,, \ p_2 = x_2 P_b$.
The kinematics of the process are fully specified by the invariants
\begin{align}
  S=(P_a+P_b)^2\,,\:s=x_1x_2S\,,\:M^2=(p_3+p_4)^2\,,\nn\\
  t_1=(p_1-p_3)^2-m_t^2\,,\: u_1=(p_1-p_4)^2-m_t^2\,.
\end{align}
It is convenient to work with two additional variables: the velocity of the
top-quarks in the \ttbar rest frame $\beta_t$ and the scattering angle $\theta$ of the
top-quark in the centre-of-mass frame of the event. These can be expressed via the invariants above as
\begin{align}
\label{eq:variables}
  \beta_t=\sqrt{1-\frac{4m_t^2}{M^2}}\,,\:\cos \theta = \frac{1}{\beta_t}\left(1+\frac{2t_1}{M^2}\right)\,.
\end{align}

\subsection{Factorisation formula for top-quark pair production}

In order to fully appreciate the ingredients appearing in the factorisation of
$0$-jettiness for heavy quark pair production, it is instructive to begin by
examining the structure of the factorisation theorem for $\Tau_0$ in the case
of colour-singlet production, which was first derived in
Ref.~\cite{Stewart:2009yx}. Defining $\tau_B\equiv \Tau_0/M$, with $M$ the
mass of the colour singlet object, 
one can write the factorisation theorem, differential in the Born phase space $\Phi_0$, as
\begin{align}
\label{eq:colsingfac}
  \frac{\de\sigma}{\de\Phi_0\de\tau_B} = M \sum_{i,j} H_{ij}(\Phi_0,\mu)\int \de t_a \de t_b B_i(t_a,z_a,\mu) B_j(t_b, z_b, \mu) S_B\left(\!\!M\tau_B-\!\frac{t_a+t_b}{M},\mu\!\right)\,,
\end{align}
where the objects appearing in this formula are UV-renormalised at a common scale $\mu$.
The differential cross section has been separated into hard, soft and beam pieces, each of which is free from large logarithms when it is evaluated at its own characteristic scale. The hard functions $H_{ij}$ arise from the matching of the effective theory onto QCD, and describes the hard dynamics of the specific process under consideration.\footnote{In order to be completely general over colour singlet processes, we have included the flux factor and other process-specific prefactors within the hard functions in our treatment.}
The beam functions $B_{i,j}$ describe collinear radiation from the initial
state partons which forms, in the language of Ref.~\cite{Stewart:2009yx}, an
`initial-state jet' with virtuality $t_{a,b}$ and light-cone momentum fraction
$z_{a,b}$. These functions are labelled by
the flavour of the incoming parton and are process-independent: their perturbative parts were
computed up to two-loop order for quarks in
Refs.~\cite{Stewart:2010qs,Gaunt:2014xga} and for gluons in
Refs.~\cite{Berger:2010xi,Gaunt:2014cfa}, and were extended to three-loop order in
Ref.~\cite{Ebert:2020unb}. Finally, the soft function $S_B$ describes the
accompanying soft radiation. It was calculated for colour-singlet processes at
two-loop order in Refs.~\cite{Kelley:2011ng,Monni:2011gb} and remains unknown at
three-loop order. The objects discussed here all have operatorial definitions in
SCET, which can be found in {\it e.g.} Refs.~\cite{Stewart:2009yx,Stewart:2010qs}. 

The analogous factorisation theorem for the \ttbar process may be derived using SCET and heavy-quark effective theory (HQET). We present its derivation in \app{Hvqfactheorem}: the final result reads

\begin{align}
  \label{eq:momspacefac}
  \frac{\de\sigma}{\de\Phi_0\de\tau_B}&=
  M \sum_{i,j}  \int \de t_a \de t_b B_i(t_a,z_a,\mu) B_j(t_b,z_b,\mu)\, \\
  &\qquad \qquad  \times \mathrm{Tr}\bigg\{\mathbf{H}_{ij}(\Phi_0,\mu)\,\mathbf{S}_{B,\,ij}\left(\!M\tau_B-\frac{t_a+t_b}{M},\Phi_0,\mu\!\right)\bigg\}\nn \, ,
\end{align}
where the sum runs over the possible channels $i,j=\{q\bar{q},\bar{q}q,gg\}$, including all light quark flavours, and $M$ is the top-quark pair invariant mass.
The charge-conjugated quark-initiated channels give equivalent contributions at leading order but they start to differ at higher orders, leading to the well-known charge asymmetry in top-quark pair production.

The main difference compared to the colour singlet production case is that the
hard and soft functions are now matrices in colour space. The soft functions
also fully depend on the LO kinematics of the process.\footnote{Following its
  operatorial definition in \eq{softoper} the soft function contains also a
  nonperturbative component, which is usually included via dedicated models in resummed calculations.  For the rest of this paper, we will only consider
the perturbative part of the soft function. After implementing the results of
this calculation fully into \geneva, these nonperturbative effects can be
recovered via the hadronisation model in the shower Monte Carlo.} The beam functions remain process-independent objects and have the same definition as in the colour singlet case.

Despite its apparent simplicity, \eq{momspacefac} is an important result of
this paper. It establishes a framework for the resummation of $\tau_B$, fully differential in the Born phase space, which is valid at any logarithmic order near the soft or collinear limits where $\tau_B \ll 1$.

\subsection{Factorisation theorem in Laplace space}
Proceeding with the evaluation of \eq{momspacefac} in momentum space is rather tedious, since the ingredients of the factorisation theorem are distribution-valued and are tied together by convolutions. In Laplace space, however, each object can be expressed in terms of logarithms of the conjugate variable and the theorem is written as a product rather than a convolution. This simplifies the renormalisation of the soft function and the consistency checks between the anomalous dimensions, and allows one to solve the evolution equations as ordinary differential equations. Once all solutions have been found, the inverse transform can be applied to the resummed pieces which are then combined to obtain the correct expression in momentum space. 

Taking the Laplace transform of \eq{momspacefac} with respect to $M\tau_B$, therefore, we find that (suppressing some of the arguments of the functions for brevity)
\begin{align}
    \mathcal{L}\left[\frac{\de\sigma}{\de\Phi_0\de\tau_B}\right]&=\\ M & \int \de(M\tau_B)  e^{-\lambda M \tau_B}\int \de t_a \de t_b B_i(t_a) B_j(t_b) \,\mathrm{Tr}\bigg \{\mathbf{H}_{ij}\,\mathbf{S}_{B,\, ij}\left(M\tau_B-\frac{t_a+t_b}{M}\right)\bigg\}\nn\\ = M\, &\int \de\Omega e^{-\lambda \Omega} \, \mathrm{Tr}\big\{\mathbf{H}_{ij}\,\mathbf{S}_{B,\,ij}(\Omega)\big\}\int \de t_a e^{-\frac{\lambda t_a}{M}} B_i(t_a) \int \de t_b e^{-\frac{\lambda t_b}{M}} B_j(t_b) \,,\nn
\end{align}
where we have defined \mbox{$\Omega\equiv M\tau_B-(t_a+t_b)/M$} and the Laplace transform is identified by $\mathcal{L}\left[\ldots\right]$.
Using the definitions of the Laplace transforms of the beam $\tilde{B} = \mathcal{L}\left[ B \right]$  and soft $\tilde{\mathbf{S}} = \mathcal{L}\left[ \mathbf{S} \right]$  functions and defining $\kappa=e^{-\gamma_E}/\lambda$, we finally arrive at
\begin{align}
\label{eq:laplacefac}
    \mathcal{L}\left[\frac{\de\sigma}{\de\Phi_0\de\tau_B}\right]&=M\, \mathrm{Tr}\bigg\{\mathbf{H}_{ij}\, \tilde{\mathbf{S}}_{B,\, ij}\left(\ln\left(\frac{1}{\lambda e^{\gamma_E} \,\mu}\right)\right)\bigg\}\,\tilde{B}_i\left(\ln\left(\frac{M}{\lambda\mu^2 e^{\gamma_E}}\right)\right)\tilde{B}_j\left(\ln\left(\frac{M}{\lambda\mu^2 e^{\gamma_E}}\right)\right) \nonumber\\
    &= M\, \mathrm{Tr}\bigg\{\mathbf{H}_{ij}\, \tilde{\mathbf{S}}_{B,\,ij}\left(\ln \frac{\kappa^2}{\mu^2} \right)\bigg\}\tilde{B}_i\left(\ln \frac{M\kappa}{\mu^2} \right)\tilde{B}_j\left(\ln \frac{M\kappa}{\mu^2} \right) \,. 
\end{align}
We have thus been able to express the differential cross section as a product of functions in Laplace space. Moreover, the Laplace-transformed soft function in \eq{laplacefac} can be written as a polynomial in the logarithm of the Laplace variable $\kappa$, with function-valued coefficients.

We are now in a position to solve the evolution equations to all orders and hence perform the resummation. We consider the various ingredients of the factorisation theorem in turn.

\subsection{The hard function and its evolution}

The colour-decomposed hard functions $\mathbf{H}_{ij}(\Phi_0,\mu)$ for
\ttbar production were first computed at one-loop order in
Ref.~\cite{Ahrens:2010zv}.
The two-loop amplitudes which are necessary for the
construction of the NNLO hard functions can instead be found in
Ref.~\cite{Chen:2017jvi}. From hereon we express the $\Phi_0$ dependence in terms of
the variables $\betat,\theta$ defined in \eq{variables} and the top-quark pair
invariant mass $M$.  Dropping the channel subscripts for ease of notation,
each hard function satisfies the following RG equation~\cite{Ahrens:2010zv}
\begin{align}
\frac{\de}{\de \ln \mu} \mathbf{H}(M,\beta_t,\theta,\mu) = \mathbf{\Gamma}_H(M,\beta_t,\theta,\mu) \, \mathbf{H}(M,\beta_t,\theta,\mu)  + \mathbf{H}(M,\beta_t,\theta,\mu) \, \mathbf{\Gamma}^\dagger_H(M,\beta_t,\theta,\mu) \, ,
\end{align}
where we conveniently wrote the anomalous dimension 
\begin{align}
\mathbf{\Gamma}_H(M,\beta_t,\theta,\mu) = \Gamma_{\mathrm{cusp}}(\alpha_s) \bigg(\ln \frac{M^2}{\mu^2} -i \pi \bigg)\, + \bm{\gamma}^h(M,\beta_t,\theta,\alpha_s)\, .
\end{align}
The non-cusp anomalous dimension matrices $\bm{\gamma}^h$ were computed up to two-loop order
in Refs.~\cite{Ferroglia:2009ep,Ferroglia:2009ii}. The all-order solution can be written as~\cite{Ahrens:2010zv}
\begin{align}
\mathbf{H}(M,\beta_t,\theta,\mu) = \mathbf{U}(M,\beta_t,\theta,\mu_h,\mu)
\mathbf{H}(M,\beta_t,\theta,\mu_h)
\mathbf{U}^\dagger(M,\beta_t,\theta,\mu_h,\mu)\,,
\end{align}
where $\mu_h$ is a hard scale of the process, {\it e.g. } the \ttbar invariant
mass $M$, such that the hard function is free from large logarithms.  When
evaluated at a generic scale $\mu$ instead of at the hard scale $\mu_h$,
the matrix $\mathbf{U}$ performs the resummation of these hard logarithms.

For later convenience, we use the fact that $\mathbf{U}$ can be rewritten by separating
out a part which comes from the cusp evolution and is diagonal in colour space
and a leftover piece $\mathbf{u}$ which also contains non-diagonal
contributions:
\begin{align}
    \mathbf{U}(M,\beta_t,\theta,\mu_h,\mu) = \exp\left[2S(\mu_h,\mu)-a_\Gamma(\mu_h,\mu)\left(\ln\frac{M^2}{\mu_h^2}-i\pi\right)\right]\mathbf{u}(M,\beta_t,\theta,\mu_h,\mu)\,.
\end{align}
The double and single logarithmic resummation are provided by the functions $S$ and $a_\Gamma$ respectively, defined as
\begin{align}
    S(\mu_a,\mu_b)&=-\int^{\alpha_s(\mu_b)}_{\alpha_s(\mu_a)}\de\alpha\frac{\Gamma_\cusp(\alpha)}{\beta(\alpha)}\int^\alpha_{\alpha_s(\mu_a)}\frac{\de\alpha'}{\beta(\alpha')},\, \nn \\
     a_\Gamma(\mu_a,\mu_b)&=-\int^{\alpha_s(\mu_b)}_{\alpha_s(\mu_a)}\de\alpha \frac{\Gamma_\cusp(\alpha)}{\beta(\alpha)} \, .
\end{align}
The off-diagonal, non-cusp evolution is instead provided by the colour matrix
\begin{align}\label{eq:upath}
    \mathbf{u}(M,\beta_t,\theta,\mu_h,\mu)=\mathcal{P}\exp\int^{\alpha_s(\mu)}_{\alpha_s(\mu_h)}\frac{\de\alpha}{\beta(\alpha)}\bm{\gamma}^h(M,\beta_t,\theta,\alpha)\,,
\end{align}
where the symbol $\mathcal{P}$ specifies the path-ordered exponential.  All of
the previous ingredients $S,a_\Gamma$ and $\mathbf{u}$ are channel-specific
and their exact definition depends on whether one is examining the quark or
gluon-initiated case. Their explicit expressions can be found in
{\it e.g.} the appendix of Ref.~\cite{Ahrens:2010zv}.

In all functions so far, we have
highlighted the dependence on both the invariant mass $M$ of the $t \bar{t}$
pair and on the variable $\beta_t$. These are related by \eq{variables}
through the value of the top-quark mass $m_t$. In order to simplify the
notation, from hereon we will drop the explicit $M$ dependence in the soft
functions and in the evolution kernels, with the understanding that these
objects still implicitly depend on $m_t$.

\subsection{The soft function and its evolution}
To the best of our knowledge, the soft function for \ttbar production which appears in \eq{momspacefac} has been defined for the first time in this work. In this section, we therefore compute the function at one-loop order, which is a necessary ingredient for resummation of the logarithms of $\Tau_0$ at NLL$^\prime$ accuracy and beyond.

\subsubsection{Calculation of the one-loop soft function}

The integrated soft functions in momentum space are given by
\begin{align}
\label{eq:softdef}
\mathbf{S}_{B,ij}(\Tau_s,\betat,\theta,\mu) = \int \de k_a^+ \de k_b^+ \, \mathbf{S}_{ij}(k_a^+,k_b^+,\betat,\theta,\mu) \, \delta(\Tau_s -k_b^+ - k_a^+)\, .
\end{align}
where the channel indices $i,j=\{q\bar{q},\bar{q}q,gg\}$. The operatorial definition in SCET is given by \eq{softoper}.
We expand the soft functions in $\alpha_s$ as
\begin{align}
\mathbf{S}_{ij}(k_a^+,k_b^+,\betat,\theta,\mu) = \mathbf{s}^{(0)}_{
  ij} \delta(k_a^+)\delta(k_b^+) + \bigg(\frac{\alpha_s}{4
  \pi}\bigg)\,   \mathbf{S}^{(1)}_{ij}(k_a^+,k_b^+,\betat,\theta,\epsilon,\mu) + \ord{\as^2},
\label{eq:softexpbare}
\end{align}
where we have expressed the bare coupling $\as^0$ in terms of the renormalised coupling $\as(\mu)$ in the $\overline{\mathrm{MS}}$ scheme using the relation $Z_{\alpha_s}\, \as(\mu) \mu^{2\epsilon}=e^{-\gamma_E\epsilon}(4\pi)^\epsilon \as^0$. The leading order (LO) coefficients $\mathbf{s}^{(0)}_{ij}$ for the $q\bar{q}$ and $gg$ channels are  defined in eq.~(65) of Ref.~\cite{Ahrens:2010zv}.
The next-to-leading order (NLO) bare soft functions in momentum space can be written  as
\begin{align}
\label{eq:softcolints}
    \mathbf{S}^{(1)}_{\mathrm{bare},\,ij}(k_a^+,k_b^+,\betat,\theta,\epsilon,\mu) = \sum_{\alpha ,\,\beta} \textbf{\textit{w}}^{\alpha\beta}_{ij}\,  \hat{\mathcal{I}}_{\alpha\beta}(k_a^+,k_b^+,\betat,\theta,\epsilon,\mu)\, ,
\end{align}
where the colour matrices $\textbf{\textit{w}}^{\alpha\beta}_{ij}$ for
the $q\bar{q}$ and $gg$ channels are defined in eq.~(71) of
Ref.~\cite{Ahrens:2010zv}  and the integrals are defined as
\begin{align}
\hat{\mathcal{I}}_{\alpha\beta}(k_a^+&,k_b^+,\betat,\theta,\epsilon,\mu) = -\frac{2 (\mu^{2}e^{\gamma_{E}})^\epsilon}{\pi^{1-\epsilon}} \,  \int \de^d k \frac{v_\alpha \cdot v_\beta }{v_\alpha\cdot k\, v_\beta\cdot k}\, \delta(k^2)\Theta(k^0) \, \\
& \times \big[ \delta(k_a^+ - k\cdot n_a)\Theta(k\cdot n_b - k\cdot n_a) \, \delta(k_b^+) + \delta(k_b^+ - k\cdot n_b)\Theta(k\cdot n_a - k\cdot n_b) \, \delta(k_a^+)\big]  \, .\nn
\end{align}
The second line of the above equation represents the observable-specific
measurement function; $v_\alpha, v_\beta =\{n_a,n_b, v_3,v_4\}$ are 
the velocities of the four external particles.
Due to the particular structure of the measurement function at NLO, we may
separate the integrals over the two hemispheres of the event, writing  
\begin{align}
 \hat{\mathcal{I}}_{\alpha\beta}(k_a^+,k_b^+,\betat,\theta,\epsilon,\mu) = \delta(k_a^+) \, \mathcal{I}_{\alpha\beta}(k_b^+,\betat,\theta,\epsilon,\mu) + \delta(k_b^+) \, \mathcal{I}_{\alpha\beta}(k_a^+,\betat,\theta,\epsilon,\mu)\, ,
\end{align}
where the integrals $\mathcal{I}_{\alpha\beta}$ refer to the contribution of a single
hemisphere.

Analogous to the decomposition for the renormalised function, it follows from \eqs{softdef}{softcolints} that
\begin{align}
\mathbf{S}^{(1)}_{B, \mathrm{bare},ij}(\Tau_s,\betat,\theta,\mu) &= \int \de k_a^+ \de k_b^+ \,  \mathbf{S}^{(1)}_{\mathrm{bare},ij}(k_a^+,k_b^+,\betat,\theta,\epsilon,\mu) \, \delta(\Tau_s - k_a^+ - k_b^+) \, \nonumber \\
&=\, 2 \sum_{\alpha,\beta} \textbf{\textit{w}}^{\alpha\beta}_{ij}\, \bigg(\int \de \tausq  \mathcal{I}_{\alpha\beta}(\tausq,\betat,\theta,\epsilon,\mu)\, \delta(\Tau_s-\tausq)\bigg)\,.
\end{align}

In the following, we therefore focus on the calculation of the single hemisphere integrals $\mathcal{I}_{\alpha\beta}(\tausq,\betat,\theta,\epsilon,\mu)$ which we normalise as
\begin{equation}
\mathcal{I}_{\alpha\beta}(\tausq,\betat,\theta,\epsilon,\mu)=-\frac{2\,  (\mu^{2}e^{\gamma_{E}})^\epsilon}{\pi^{1-\epsilon}}I_{\alpha\beta}(\tausq,\betat,\theta,\epsilon,\mu) \,.
\end{equation}

The symmetries of the integrals mean that we have the following relations:
\begin{align}
    I_{v_4 v_4}&=I_{v_3 v_3}(\theta\rightarrow  \pi - \theta)\nonumber \\
    I_{n_av_4}&=I_{n_av_3}(\theta\rightarrow \pi - \theta) \nonumber \\
    I_{n_bv_4}&=I_{n_b v_3}(\theta\rightarrow \pi - \theta)
\end{align}

The integrals which feature only light-like legs are also needed in the
computation of the $0$-jettiness soft function for colour singlet production,
and as such appear in \textit{e.g.}
Refs.~\cite{Jouttenus:2011wh,Kelley:2011ng,Monni:2011gb}. In \app{softints}, we evaluate
analytically the pole structure of all remaining required integrals $I_{\alpha\beta}$ at this order. We were unable to obtain closed analytic expressions for the finite parts, however, and have therefore evaluated these onefold integrals numerically.

Thus far we have discussed the computation of the bare soft function at $\ord{\as}$. This object still contains ultraviolet poles in $\epsilon$ and must therefore be renormalised. In Laplace space, the renormalisation can be performed by multiplying both sides of the bare soft function by a matrix-valued renormalisation factor
\begin{align}
  \tilde{\mathbf{S}}_{B,ij}(L,\betat,\theta,\mu)=\mathbf{Z}^\dagger_{S,ij}(\as,L,\epsilon)\, \tilde{\mathbf{S}}_{B, \mathrm{bare},ij}(L,\betat,\theta)\,\mathbf{Z}_{S,ij}(\as,L,\epsilon)\,,
  \label{eq:softrenorm}
\end{align}
where we have defined $L=\ln (\kappa^2/\mu^2)$. At one-loop order, the effect of \eq{softrenorm} is simply to remove the UV poles from $\tilde{\mathbf{S}}^{(1)}_{B, _{\mathrm{bare}}}$. The renormalised soft function at this order is therefore obtained by taking the finite part of the bare function. Since the bare soft function in \eq{softrenorm} is independent of $\mu$, both the renormalised function and the renormalisation factor obey a renormalisation group equation, which we shall discuss in the following section.

The renormalisation procedure also completely determines the structure of the $\ord{\as}$ term $\mathbf{Z}_S^{(1)}$, which allows us to extract the soft anomalous dimension at one-loop.  We verified that by doing so, this object satisfies consistency relations required by RG invariance of \eq{momspacefac} (see \eq{consistency}). In addition, by exploiting this relation at one order higher, we are able to extract the soft anomalous dimension at two-loop order.

\subsubsection{Solving the soft RG equations at fixed order}
\label{sec:2loopsoft}
A resummation at full NNLL$^\prime$ accuracy would require knowledge of the
two-loop contributions to the soft function, which have not yet been
calculated. It is, however, possible to obtain partial knowledge about the
two-loop function by solving the renormalisation group evolution equations at
fixed order. In this way, one can obtain the logarithmic terms at
$\ord{\alpha_s^2}$ expressed in terms of coefficients at lower order, leaving
only the term proportional to $\delta(\Tau_0)$ to be determined by an explicit calculation.

The soft functions in Laplace space satisfy the following renormalisation group equations
\begin{align}\label{eq:softRG}
\frac{\de}{\de \ln \mu} \tilde{\mathbf{S}}_{B}(L,\beta_t,\theta,\mu) = \bigg[\Gamma_\cusp L\, - \bm{\gamma}^{s^\dagger}\bigg]\, \tilde{\mathbf{S}}_{B}(L,\beta_t,\theta,\mu) + \tilde{\mathbf{S}}_{B}(L,\beta_t,\theta,\mu) \,\bigg[\Gamma_\cusp L\, - \bm{\gamma}^{s}\bigg] \, ,
\end{align}
where we have dropped the channel
subscript for simplicity.
Since the expansions of $\Gamma_\cusp$ and the non-cusp soft anomalous
dimension matrices $\bm{\gamma}^{s}$ start at $\ord{\as}$, defining
\begin{align}
\tilde{\mathbf{S}}_{B}(L,\beta_t,\theta,\mu) = \mathbf{s}^{(0)} + \frac{\alpha_s}{4 \pi} \tilde{\mathbf{S}}^{(1)}_{B}+  \bigg( \frac{\alpha_s}{4 \pi} \bigg)^2 \tilde{\mathbf{S}}^{(2)}_{B} + \mathcal{O}(\alpha^3_s)
\label{eq:softexp}
\end{align}
 and expanding \eq{softRG} at NNLO we have
\begin{align}
    \frac{\de}{\de L} \tilde{\mathbf{S}}_{B}^{(2)} = \frac{1}{2}\tilde{\mathbf{S}}^{(1)}_{B}\bigg[(-\Gamma_\cusp^{(0)}L - \beta_0) + \bm{\gamma}^{s(0)} \bigg]
    + \frac{1}{2} \mathbf{s}^{(0)}\bigg[-\Gamma_\cusp^{(1)} L + \bm{\gamma}^{s(1)} \bigg] +\, \mathrm{h.c.}
\end{align}
Denoting further the logarithmic coefficients of the soft function as
\begin{align}
\tilde{\mathbf{S}}_{B}(L,\beta_t,\theta,\mu) = \sum_{n=0}^\infty \,\sum_{m=0}^{2n} \left(\frac{\as}{4\pi}\right)^n\, \tilde{\mathbf{S}}^{(n,m)}_{B}(\beta_t,\theta) L^m
\end{align}
and again suppressing arguments for brevity, we find the solution
\begin{align}
  \label{eq:softcoeffs}
    \tilde{\mathbf{S}}^{(2,4)}_{B} &= -\frac{1}{8}\tilde{\mathbf{S}}^{(1,2)}_{B}\Gamma_\cusp^{(0)} +\, \mathrm{h.c.}\nn \\
    \tilde{\mathbf{S}}^{(2,3)}_{B} &= \frac{1}{6}\left(-\tilde{\mathbf{S}}^{(1,1)}_{B}\Gamma_\cusp^{(0)} + \tilde{\mathbf{S}}^{(1,2)}_{B}\bm{\gamma}^{s(0)} - \beta_0\tilde{\mathbf{S}}^{(1,2)}_{B} \right)+\, \mathrm{h.c.} \\
    \tilde{\mathbf{S}}^{(2,2)}_{B} &= \frac{1}{4}\left(-\tilde{\mathbf{S}}^{(1,0)}_{B}\Gamma_\cusp^{(0)} + \tilde{\mathbf{S}}^{(1,1)}_{B}\bm{\gamma}^{s(0)} -\mathbf{s}^{(0)}\Gamma_\cusp^{(1)} - \beta_0\tilde{\mathbf{S}}^{(1,1)}_{B} \right)+\, \mathrm{h.c.}\nn \\
    \tilde{\mathbf{S}}^{(2,1)}_{B} &= \frac{1}{2}\left(\tilde{\mathbf{S}}^{(1,0)}_{B}\bm{\gamma}^{s(0)} + \mathbf{s}^{(0)}\bm{\gamma}^{s(1)} -\beta_0\tilde{\mathbf{S}}^{(1,0)}_{B}\right) +\, \mathrm{h.c.}\nn 
\end{align}

Upon transforming back to momentum space, we thus have all the soft
ingredients necessary to construct the $\Tau_0$ spectrum at approximate NNLO.
We are only missing the term $\tilde{\mathbf{S}}^{(2,0)}_{B}$, which
contributes only at the point $\Tau_0=0$ and must be computed separately. 
This means that once we combine these with the contributions coming from the
beam and hard functions we are able to cancel all the singular pieces at small
$\Tau_0$ of the NLO calculation for \ttbarj production.

\subsubsection{Evolution}

In Laplace space, the all-order solutions of the soft RG evolution in \eq{softRG} can be written as
\begin{align}
\tilde{\mathbf{S}}_{B}(L,\beta_t,\theta,\mu) = \mathbf{V}^{\dagger}(\kappa,\beta_t,\theta,\mu_s,\mu)\, \tilde{\mathbf{S}}_{B}(L,\beta_t,\theta,\mu_s)\, \mathbf{V}(\kappa,\beta_t,\theta,\mu_s,\mu) \, ,
 \label{eq:softevol}
\end{align}
where the unitary matrix $\mathbf{V}$ satisfies the differential equation
\begin{align}
\frac{\de}{\de \ln \mu}\mathbf{V}(\kappa,\beta_t,\theta,\mu_s,\mu) = \bigg(\Gamma_{\mathrm{cusp}}\, \ln \frac{\kappa^2}{\mu^2} - \bm{\gamma}_s \bigg)\, \mathbf{V}(\kappa,\beta_t,\theta,\mu_s,\mu) \,, 
\end{align}
and the soft scale $\mu_s \sim \Tau_0$ minimises the logarithms in the soft functions. 
Proceeding analogously to the hard function case and resumming the soft
logarithms while evolving from the soft scale to a generic scale $\mu$, we find the solution
\begin{align}
    \mathbf{V}(\kappa,\beta_t,\theta,\mu_s,\mu) = \exp\left[2S(\mu_s,\mu)\right]\left(\frac{\kappa^2}{\mu_s^2}\right)^{-a_\Gamma(\mu_s,\mu)}\mathbf{v}(\beta_t,\theta,\mu_s,\mu),
\end{align}
with the non-cusp soft evolution matrices given by
\begin{align}
    \mathbf{v}(\beta_t,\theta,\mu_s,\mu)=\mathcal{P}\exp\left\{-\int^{\alpha_s(\mu)}_{\alpha_s(\mu_s)}\frac{\de\alpha}{\beta(\alpha)}\bm{\gamma}^{s}(\beta_t,\theta,\alpha)\right\}.
\end{align}
Substituting these ingredients into \eq{softevol} we obtain
\begin{align}
  \tilde{\mathbf{S}}_{B}(L,\beta_t,\theta,\mu) = \exp\left[4S(\mu_s,\mu)\right]\mathbf{v}^{\dagger}(\beta_t,\theta,\mu_s,\mu)\, \tilde{\mathbf{S}}_{B}(\partial_{\eta_s},\beta_t,\theta,\mu_s)\,\mathbf{v}(\beta_t,\theta,\mu_s,\mu)\left(\frac{\kappa^2}{\mu_s^2}\right)^{\eta_s}
\end{align}
where $\eta_s\equiv-2a_\Gamma(\mu_s,\mu)$.  In the last equation we have
rewritten the logarithms appearing as an argument of the soft function in
terms of partial derivatives acting on the last factor~\cite{Becher:2006nr,Becher:2014oda}.
Transforming back to momentum space yields
\begin{align}\label{eq:softRGsol}
  \mathbf{S}_{B}(\tausq,\beta_t,\theta,\mu) =&
  \exp\left[4S(\mu_s,\mu)\right]\mathbf{v}^{\dagger}(\beta_t,\theta,\mu_s,\mu)\,
  \tilde{\mathbf{S}}_{B}(\partial_{\eta_s},\beta_t,\theta,\mu_s)\,\mathbf{v}(\beta_t,\theta,\mu_s,\mu)
  \nn \\ & \times \frac{1}{\tausq}\left(\frac{\tausq}{\mu_s}\right)^{2\eta_s} \, \frac{e^{-2\gamma_E\eta_s}}{\Gamma(2\eta_s)}\, .
\end{align}
Due to the RG invariance of the full cross section we have the following relation between the non-cusp anomalous dimensions of the hard, soft, and beam functions
\begin{align}\label{eq:consistency}
\bm{\gamma}^s = \bm{\gamma}^h + \gamma^B\, \mathbf{1} \, ,
\end{align}
where the non-diagonal part of the soft anomalous dimension arises entirely
from the non-cusp anomalous dimension of the hard function and $\gamma^B$ is the
non-cusp anomalous dimension for the beam function.
The evolution formula in \eq{softRGsol} for the soft function can therefore be rewritten as
\begin{align}\label{eq:softRGsol2}
  \mathbf{S}_{B}(\tausq,\beta_t,\theta,\mu) = &\exp\left[4S(\mu_s,\mu) + 2 a_{\gamma^B}(\mu_s,\mu)\right]\,\\
  &\times \, \mathbf{u}^{\dagger}(\beta_t,\theta,\mu,\mu_s)\, \tilde{\mathbf{S}}_{B}(\partial_{\eta_s},\beta_t,\theta,\mu_s)\,\mathbf{u}(\beta_t,\theta,\mu,\mu_s) \, \frac{1}{\tausq}\left(\frac{\tausq}{\mu_s}\right)^{2\eta_s} \, \frac{e^{-2\gamma_E\eta_s}}{\Gamma(2\eta_s)}\, ,\nn
\end{align}
where the order of the scale arguments in the $\mathbf{u}$ evolution matrices is now inverted relative to the $\mathbf{v}$ matrices and
\begin{align}
a_{\gamma^B}(\mu_s,\mu) = - \int_{\alpha_s(\mu_s)}^{\alpha_s(\mu)}\, \de \alpha \frac{\gamma^B(\alpha)}{\beta(\alpha)} \, .
\end{align}

\subsection{The beam functions and their evolution}

The process-independent $\Tau_0$ beam functions $B_i$  have been computed up
to N$^3$LO accuracy and are available in the
literature~\cite{Stewart:2010qs,Gaunt:2014xga,Berger:2010xi,Gaunt:2014cfa,Ebert:2020unb}. The quark and gluon beam functions satisfy the following RG equation in Laplace space
\begin{align}
\frac{\de}{\de \ln \mu} \tilde{B}_i(L_c,z,\mu) = \bigg[ - 2\, \Gamma_{\mathrm{cusp}}(\alpha_s)\, L_c + \gamma^{B}_i(\alpha_s)\bigg] \tilde{B}_i(L_c,z,\mu)\, ,
\end{align}
where the index $i=\{q,\bar{q},g\}$,  $L_c=\ln\big[ (M \kappa)/\mu^2\big]$ and $\Gamma_{\mathrm{cusp}}=C_D
\gamma_{\mathrm{cusp}}$ with $C_D=\{C_F,C_A\}$ for the quark and the gluon
beam functions respectively.
The explicit expressions for the non-cusp beam anomalous dimensions 
$\gamma^B_i$ up to NNLO can be found in {\it e.g.} Appendix D of Ref~\cite{Billis:2019vxg}.  
Dropping the flavour index for brevity, the evolution equation has the solution
\begin{align}
    \tilde{B}(L_c,z,\mu)=\exp\left[-4S(\mu_B,\mu)-a_{\gamma^B}(\mu_B,\mu)\right] \tilde{B}(\partial_{\eta_B},z,\mu_B)\left(\frac{
    M\kappa}{\mu_B^2}\right)^{\eta_B}\,,
\end{align}
where $\eta_B\equiv 2a_\Gamma(\mu_B,\mu)$ and $\mu_B \sim  \sqrt{\phantom{|}\!\! \Tau_0 M}$ is
the beam scale. Taking the inverse transform again we find that, in momentum space,
\begin{align}
    B(t,z,\mu)=\exp\left[-4S(\mu_B,\mu)-a_{\gamma^B}(\mu_B,\mu)\right]
    \tilde{B}(\partial_{\eta_B},z,\mu_B)\frac{1}{t}\left(\frac{t}{\mu_B^2}\right)^{\eta_B}
    \, \frac{e^{-\gamma_E\eta_B}}{\Gamma(\eta_B)} \,.
\end{align}

\section{Resummation via renormalisation group evolution}
\label{sec:resummation}
In this section, we combine the factorisation theorem and the perturbative ingredients presented in \sec{FacTau} to resum logarithms of $\Tau_0/M$. We present explicit formul\ae~for the resummed $\Tau_0$ spectrum at NLL$^\prime$, NNLL and NNLL$^\prime$ order.  
\subsection{All-order solutions of the RG equations}

Substituting the resummed expressions for the ingredients of  \eq{momspacefac}
which we have presented in \sec{FacTau} and after integrating over the
virtualities $t_a$ and $t_b$, we are able to write the resummed cross section
in a compact form as
\begin{align}
\frac{\de\sigma}{\de\Phi_0\de\tau_B}& = U(\mu_h,\mu_B,\mu_s,L_h,L_s) \nn\\& \times \mathrm{Tr}\bigg\{\mathbf{u}(\beta_t,\theta,\mu_h,\mu_s)\, \mathbf{H}(M,\beta_t,\theta,\mu_h)\,\mathbf{u}^{\dagger}(\beta_t,\theta,\mu_h,\mu_s)\,\tilde{\mathbf{S}}_{B}(\partial_{\eta_s}+L_s,\beta_t,\theta,\mu_s)\bigg\}\nn\\ &\times  \tilde{B}_a(\partial_{\eta_B} + L_B,z_a,\mu_B) \tilde{B}_b(\partial_{\eta_B'} + L_B,z_b,\mu_B) \frac{1}{\tau_B^{1-\eta_{\mathrm{tot}}}}\frac{e^{-\gamma_E\eta_{\mathrm{tot}}}}{\Gamma(\eta_{\mathrm{tot}})}\,.
\label{eq:finalformula}
\end{align}
The derivative terms inside the arguments of the soft and beam functions act on the factor in the last line of the previous
equation, which we refer to as the generating function. In the previous formula we have defined
\begin{align}
\label{eq:diagonalU}
U(\mu_h,&\mu_B,\mu_s,L_h,L_s)  = \\& \exp\bigg[4S(\mu_h,\mu_B)+4S(\mu_s,\mu_B)+2a_{\gamma^B}(\mu_s,\mu_B)-2a_\Gamma(\mu_h,\mu_B)\,L_h -2a_\Gamma(\mu_s,\mu_B)\,L_s\bigg]\,.\nn
\end{align}
We have also introduced the quantities $\eta_s\equiv2a_\Gamma(\mu,\mu_s)$, $\eta_B\equiv
2a_\Gamma(\mu_B,\mu)$, $\eta_{\mathrm{tot}}=2\eta_s+\eta_B+\eta_B'$, and we
explicitly write the beam, soft and hard logarithms as  $L_B=\log (M^2/\mu_B^2)$, $L_s=\log (M^2/\mu_s^2)$ and $L_h=\log (M^2/\mu_h^2)$. For the derivation of the formula above we have used the relations
\begin{align}\label{eq:Srelations}
\mathbf{u}(\beta_t,\theta,\mu_c,\mu_a)\, \mathbf{u}(\beta_t,\theta,\mu_b,\mu_c) & = \mathbf{u}(\beta_t,\theta,\mu_b,\mu_a)\, ,\nonumber\\
    a_{\Gamma}(\mu_a,\mu_c) &= a_{\Gamma}(\mu_a,\mu_b) + a_{\Gamma}(\mu_b,\mu_c)\, ,\nonumber \\
    a_{\gamma^i}(\mu_a,\mu_c) &= a_{\gamma^i}(\mu_a,\mu_b) + a_{\gamma^i}(\mu_b,\mu_c) \, , \nonumber \\
    S(\mu_a,\mu_b)-S(\mu_c,\mu_b) &= S(\mu_a,\mu_c) - a_{\Gamma}(\mu_c,\mu_b)\log\frac{\mu_a}{\mu_c}\, .
\end{align}
to simplify the final expressions.

The expression in \eq{finalformula} is our master formula and the primary
outcome of this work. It is formally valid at all logarithmic orders. It is
possible to evaluate it at NLL$^\prime$, NNLL and NNLL$^\prime$ depending on the
order in $\alpha_s$ at which the anomalous dimensions and the boundary terms are available.

In order to evaluate $\mathbf{u}$ we first find the matrix $\bm{\Lambda}$ which diagonalises
the LO non-cusp hard anomalous dimension
\begin{align}
\bm{\gamma}^{h(0)}_D=\bm{\Lambda}^{-1}\bm{\gamma}^{h(0)}\bm{\Lambda}
\end{align}
and define the vector $\vec{\gamma}^{h(0)}$ consisting of the eigenvalues of
the diagonal matrix $\bm{\gamma}^{h(0)}_D$. The solution of the non-cusp
evolution matrix in \eq{upath} up to NNLL can then be obtained perturbatively as an expansion in
$\alpha_s$ following App. A of Ref.~\cite{Ahrens:2010zv} and the references
therein~\cite{Buras:1991jm,Buchalla:1995vs}. We find 
\begin{align}\label{eq:unondiag}
    \mathbf{u}^{\text{NNLL}}(\beta_t,\theta,\mu_h,\mu)= \left[ \bm{\Lambda}\left(1+\frac{\alpha_s(\mu)}{4\pi}\bm{K}\right)\left(\left[\frac{\alpha_s(\mu_h)}{\alpha_s(\mu)}\right]^{\frac{\vec{\gamma}^{h(0)}}{2\beta_0}}\right)_D\!\!\!\left(1-\frac{\alpha_s(\mu_h)}{4\pi}\bm{K}\right)\bm{\Lambda}^{-1} \right]_{\ord{\as}}
\end{align}
where $[\ldots]_{\ord{\as}}$ indicates the expansion up to the first power of $\as$ and
the elements of the matrix $\bm{K}$ are given by
\begin{align}
    K_{ij} = \delta_{ij}\vec{\gamma}^{h(0)}_i\frac{\beta_1}{2\beta_0^2}-\frac{[\bm{\Lambda}^{-1}\bm{\gamma}^{h(1)}\bm{\Lambda}]_{ij}}{2\beta_0+\vec{\gamma}^{h(0)}_i-\vec{\gamma}^{h(0)}_j}.
\end{align}

\begin{table}[t]
\centering
\begin{tabular}{c| c c c c}
\hline\hline
Accuracy & b.c. & $\beta,\,\Gamma_{\rm{cusp}}$ & $\gamma^x$ & n.s. matching\\
\hline
$\rm{NLL}^\prime$ & $\ord{\as}$ & 2-loop & 1-loop & LO$_1$\\
$\rm{NNLL}$ & $\ord{\as}$ & 3-loop & 2-loop & LO$_1$\\
$\rm{NNLL}^\prime$ & $\ord{\as^2}$ & 3-loop & 2-loop & NLO$_1$\\
\end{tabular}
\caption{Definition of resummed accuracies in terms of the required perturbative order of the resummation ingredients and of the boundary conditions. The appropriate fixed order matching is also indicated. }
\label{tab:resummedacc}
\end{table}

For later convenience we separate the contributions of the NLL evolution from
the pure NNLL corrections which enter at $\mathcal{O}(\alpha_s)$ in the
logarithmic counting. In practice we define the quantities
\begin{align}
    \mathbf{u}^{\text{NLL}}(\beta_t,\theta,\mu_h,\mu)&=\bm{\Lambda}\left(\left[\frac{\alpha_s(\mu_h)}{\alpha_s(\mu)}\right]^{\frac{\vec{\gamma}^{h(0)}}{2\beta_0}}\right)_D\bm{\Lambda}^{-1}
    \,\quad  \textrm{and}\\
    \mathbf{u}^{\text{NNLL}_{\alpha_s}}(\beta_t,\theta,\mu_h,\mu) & =  \bigg(\frac{\alpha_s(\mu)}{4\pi}\bigg)\, \bigg[\big(\bm{\Lambda}\bm{K}\, \bm{\Lambda}^{-1}\big), \mathbf{u}^{\text{NLL}}(\beta_t,\theta,\mu_h,\mu)\bigg]\, ,   
\end{align}
(where $[x,y]$ denotes the usual commutator) such that the NNLL non-cusp evolution matrix is written as
\begin{align}
   \mathbf{u}^{\text{NNLL}}(\beta_t,\theta,\mu_h,\mu)= \mathbf{u}^{\text{NLL}}(\beta_t,\theta,\mu_h,\mu) + \mathbf{u}^{\text{NNLL}_{\alpha_s}}(\beta_t,\theta,\mu_h,\mu). 
\end{align}

\subsection{NLL$^\prime$ and NNLL formul\ae}
\label{sec:nllandnnll}
Beginning from the general resummation formula in \eq{finalformula}, it is
possible to write the formula one would obtain at a given resummed accuracy
explicitly. To this end, we use the expansion of the soft function defined in \eq{softexp} and define the following perturbative coefficients for
the hard and beam  functions 
\begin{align}
\mathbf{H}(M,\beta_t,\theta,\mu_h) = \frac{\alpha^2_s}{d_R}\, \bigg[\mathbf{H}^{(0)} + \frac{\alpha_s}{4 \pi} \mathbf{H}^{(1)}+  \bigg( \frac{\alpha_s}{4 \pi} \bigg)^2 \mathbf{H}^{(2)} + \mathcal{O}(\alpha^3_s)\bigg]\, ,
\end{align}
\begin{align}
\tilde{B}_i(L_c,x,\mu_B) = \tilde{B}_i^{(0)} + \frac{\alpha_s}{4 \pi} \tilde{B}_i^{(1)}+  \bigg( \frac{\alpha_s}{4 \pi} \bigg)^2 \tilde{B}_i^{(2)} + \mathcal{O}(\alpha^3_s)\, ,
\end{align}
 with $d_R$ being the dimension of the quark or the gluon SU($N_c$) colour
representation such that $d_R=\{N_c, N_c^2-1\}$ for the $q\bar{q}$ and $gg$
channels respectively.  In \tab{resummedacc} we summarise the perturbative order to which the anomalous dimensions and beta function must be evaluated in order to reach a given accuracy.

The explicit expression at NLL$^\prime$ then reads
\begin{align} 
 \frac{\de\sigma^{\text{NLL}^\prime}}{\de\Phi_0\de\tau_B}  = &\
U^{\mathrm{NLL}}(\mu_h,\mu_B,\mu_s,L_h,L_s) \nn\\&\times\Bigg\{  \, \Bigg[\, \mathrm{Tr}\bigg\{\mathbf{u}^{\text{NLL}}(\beta_t,\theta,\mu_h,\mu_s)\, \mathbf{H}^{(0)}(M,\beta_t,\theta,\mu_h)\,\mathbf{u}^{\dagger\, \text{NLL}}(\beta_t,\theta,\mu_h,\mu_s)\,\mathbf{s}^{(0)}\, \bigg\}  \nn \, \\
& + \frac{\alpha_s(\mu_h)}{4 \pi} \, \mathrm{Tr}\bigg\{\mathbf{u}^{\text{NLL}}(\beta_t,\theta,\mu_h,\mu_s)\, \mathbf{H}^{(1)}(M,\beta_t,\theta,\mu_h)\,\mathbf{u}^{\dagger\, \text{NLL}}(\beta_t,\theta,\mu_h,\mu_s)\,\mathbf{s}^{(0)} \bigg\}\nn\\
& + \frac{\alpha_s(\mu_s)}{4 \pi}\,  \mathrm{Tr}\bigg\{\mathbf{u}^{\text{NLL}}(\beta_t,\theta,\mu_h,\mu_s)\, \mathbf{H}^{(0)}(M,\beta_t,\theta,\mu_h)\,\mathbf{u}^{\dagger\, \text{NLL}}(\beta_t,\theta,\mu_h,\mu_s)\,\nn \\
&\quad  \times \, \tilde{\mathbf{S}}^{(1)}_{B}(\partial_{\eta_s}+L_s,\beta_t,\theta,\mu_s)\bigg\}\, \Bigg]\, \times \tilde{B}^{(0)}_a(z_a,\mu_B) \, \tilde{B}^{(0)}_b(z_b,\mu_B) \, \nonumber \\
& + \frac{\alpha_s(\mu_B)}{4 \pi}  \, \mathrm{Tr}\bigg\{\mathbf{u}^{\text{NLL}}(\beta_t,\theta,\mu_h,\mu_s)\, \mathbf{H}^{(0)}(M,\beta_t,\theta,\mu_h)\,\mathbf{u}^{\dagger\, \text{NLL}}(\beta_t,\theta,\mu_h,\mu_s)\,\mathbf{s}^{(0)}\bigg\}\nn\\ &\times \bigg(\tilde{B}^{(1)}_a(\partial_{\eta_B} + L_B,z_a,\mu_B) \, \tilde{B}^{(0)}_b(z_b,\mu_B) + \tilde{B}^{(0)}_a(z_a,\mu_B) \, \tilde{B}^{(1)}_b(\partial_{\eta_B'} + L_B,z_b,\mu_B)\bigg)\, \Bigg\}\, \nonumber \\
&\times \frac{1}{\tau_B^{1-\eta^{\text{NLL}}_{\mathrm{tot}}}}\frac{e^{-\gamma_E\eta^{\text{NLL}}_{\mathrm{tot}}}}{\Gamma(\eta^{\text{NLL}}_{\mathrm{tot}})}\, ,\label{eq:nllformula}
\end{align}
where the factor $U^{\mathrm{NLL}}$ is given by \eq{diagonalU} with the
expressions for $S$, $a_{\Gamma}$ and $a_{\gamma^B}$ expanded only up to NLL accuracy.

We now consider accuracies beyond NLL$^\prime$. We write the additional $\ord{\alpha_s}$  correction that must be added to the NLL$^\prime$ result in \eq{nllformula} to reach NNLL accuracy as
\begin{align}
\frac{\de\sigma^{\text{NNLL}}}{\de\Phi_0\de\tau_B} = \frac{\de\sigma^{\text{NLL}^\prime}}{\de\Phi_0\de\tau_B} + \frac{\de\sigma^{\text{NNLL}_{\alpha_s}}}{\de\Phi_0\de\tau_B}\, .
\label{eq:resummedbreakdown1}
\end{align}

Given that the non-diagonal evolution term in \eq{unondiag} is only known
perturbatively, at NNLL we also need to expand the exponential
prefactor $U$ which performs the diagonal  evolution to order $\alpha_s$ to maintain
consistency with the perturbative expansion of the non-diagonal evolution
term.\footnote{We are always considering the logarithmic counting where $\alpha_s L \sim 1$. Note that for the $a_\Gamma$ contributions the expansion must be performed to one order higher.}

The $\ord{\alpha_s}$ contribution that ensures NNLL accuracy is then given by
\begin{align}
\frac{\de\sigma^{\text{NNLL}_{\alpha_s}}}{\de\Phi_0\de\tau_B}  =&\  U^{\mathrm{NLL}}(\mu_h,\mu_B,\mu_s,L_h,L_s) \frac{\eta^{\text{NLL}}_{\mathrm{tot}}}{\tau_B^{1-\eta^{\text{NLL}}_{\mathrm{tot}}}}\frac{e^{-\gamma_E\eta^{\text{NLL}}_{\mathrm{tot}}}}{\Gamma(1 + \eta^{\text{NLL}}_{\mathrm{tot}})} \nn \\
&\times \Bigg\{ \bigg[ \, \bigg(4S^{(\alpha_s)}(\mu_h,\mu_B)+4S^{(\alpha_s)}(\mu_s,\mu_B)  + 2a^{(\alpha_s)}_{\gamma^B}(\mu_s,\mu_B)\nn \nonumber\\
&\, -2a^{(\alpha^2_s)}_\Gamma(\mu_h,\mu_B)\,L_h \, - 2a^{(\alpha^2_s)}_\Gamma(\mu_s,\mu_B)\,L_s  + \eta^{(\alpha^2_s)}_{\mathrm{tot}} \big(\ln(\tau_B) - \gamma_E - \psi_0(\eta^{\text{NLL}}_{\mathrm{tot}})\big)\bigg)\, \nn \\
& \times \mathrm{Tr}\bigg\{\mathbf{u}^{\text{NLL}}(\beta_t,\theta,\mu_h,\mu_s)\, \mathbf{H}^{(0)}(M,\beta_t,\theta,\mu_h)\,\mathbf{u}^{\dagger\, \text{NLL}}(\beta_t,\theta,\mu_h,\mu_s)\,\mathbf{s}^{(0)}\, \bigg\}\bigg] \,  \nn \\
& +\,  \mathrm{Tr}\bigg\{\mathbf{u}^{\text{NNLL}_{\alpha_s}}(\beta_t,\theta,\mu_h,\mu_s)\, \mathbf{H}^{(0)}(M,\beta_t,\theta,\mu_h)\,\mathbf{u}^{\dagger\, \text{NLL}}(\beta_t,\theta,\mu_h,\mu_s)\,\mathbf{s}^{(0)}\, \bigg\} \,\nn \\
& +\,  \mathrm{Tr}\bigg\{\mathbf{u}^{\text{NLL}}(\beta_t,\theta,\mu_h,\mu_s)\, \mathbf{H}^{(0)}(M,\beta_t,\theta,\mu_h)\,\mathbf{u}^{\dagger\, \text{NNLL}_{\alpha_s}}(\beta_t,\theta,\mu_h,\mu_s)\,\mathbf{s}^{(0)}\, \bigg\} \,\, \Bigg\} \, \nn \\
&\times \,   \tilde{B}^{(0)}_a(z_a,\mu_B) \, \tilde{B}^{(0)}_b(z_b,\mu_B) \, .
\label{eq:NNLLcorr}
\end{align}
In the equation above the superscripts on the $S$, $a_{\gamma^B}$ and
$a_\Gamma$ functions indicate the correction at a specific order in
$\alpha_s$.\footnote{ Note that, for simplicity, we have not factored
  out any couplings from these terms. To give a specific example,
  $S^{(\as)}$ is given by the coefficient of $\as/4\pi$ of eq. A.3 in
  Ref.~\cite{Ahrens:2010zv}.}  We have also consistently expanded the
generating function at the appropriate order. Since after the
expansion no derivatives appear in the arguments of the soft and beam
functions, we are free to move the generating function to the first
line.

An alternative to \eq{NNLLcorr}, which formally still gives the same NNLL
accuracy, involves upgrading its first line to use the
NNLL formul\ae~ for the diagonal evolution factor $U$ and for the
generating function. One must then drop the expansion appearing inside
the parentheses in the second and third line.  We will discuss the numerical impact of 
this difference in \sec{resumresults}.

\subsection{NNLL$^\prime$ formul\ae}

Finally, we examine the necessary ingredients for NNLL$^\prime$ accuracy. As
discussed in \sec{2loopsoft}, we obtain all the logarithmic contributions
of the soft functions at $\mathcal{O}(\alpha^2_s)$ by RG evolution. The
two-loop hard functions and the $\delta(\mathcal{T}_0)$ contributions of the
soft functions at $\mathcal{O}(\alpha^2_s)$ must instead be calculated explicitly. 

Writing the additional $\ord{\as^2}$ term needed to reach full NNLL$^\prime$ accuracy as 
\begin{align}
\frac{\de\sigma^{\text{NNLL}^\prime}}{\de\Phi_0\de\tau_B} = \frac{\de\sigma^{\text{NNLL}}}{\de\Phi_0\de\tau_B}
+ \frac{\de\sigma^{\text{NNLL}_{\alpha^2_s}}}{\de\Phi_0\de\tau_B}\, ,
\label{eq:resummedbreakdown2}
\end{align}
we find
\begin{align}
  \label{eq:NNLLcorras2}
& \frac{\de\sigma^{\text{NNLL}_{\alpha^2_s}}}{\de\Phi_0\de\tau_B}  = U^{\mathrm{NLL}}(\mu_h,\mu_B,\mu_s,L_h,L_s)\, \nn \\
&\qquad \qquad \qquad \times \Bigg\{\, \Bigg[
{\frac{\alpha^2_s(\mu_h)}{16 \pi^2}\,  \mathrm{Tr}\bigg\{\mathbf{u}^{\text{NLL}}(\beta_t,\theta,\mu_h,\mu_s)\, \mathbf{H}^{(2)}(M,\beta_t,\theta,\mu_h)\,\mathbf{u}^{\dagger\, \text{NLL}}(\beta_t,\theta,\mu_h,\mu_s)\,\mathbf{s}^{(0)}\, \bigg\}} \, \nn \\
& + \frac{\alpha^2_s(\mu_s)}{16 \pi^2} \mathrm{Tr}\bigg\{\mathbf{u}^{\text{NLL}}(\beta_t,\theta,\mu_h,\mu_s)\, \mathbf{H}^{(0)}(M,\beta_t,\theta,\mu_h)\,\mathbf{u}^{\dagger\, \text{NLL}}(\beta_t,\theta,\mu_h,\mu_s)\,\tilde{\mathbf{S}}^{(2)}_{B}(\partial_{\eta_s}+L_s,\beta_t,\theta,\mu_s)\, \bigg\} \, \nn \\
& + \frac{\alpha_s(\mu_h)\alpha_s(\mu_s)}{16 \pi^2} \mathrm{Tr}\bigg\{\mathbf{u}^{\text{NLL}}(\beta_t,\theta,\mu_h,\mu_s)\, \mathbf{H}^{(1)}(M,\beta_t,\theta,\mu_h)\,\mathbf{u}^{\dagger\, \text{NLL}}(\beta_t,\theta,\mu_h,\mu_s)\,\nn \\
& \qquad \qquad \quad \times \, \tilde{\mathbf{S}}^{(1)}_{B}(\partial_{\eta_s}+L_s,\beta_t,\theta,\mu_s)\, \bigg\} \Bigg]\, \times  \tilde{B}^{(0)}_a(z_a,\mu_B) \, \tilde{B}^{(0)}_b(z_b,\mu_B) \, \nn \\
& + \frac{\alpha^2_s(\mu_B)}{16 \pi^2}\,  \mathrm{Tr}\bigg\{\mathbf{u}^{\text{NLL}}(\beta_t,\theta,\mu_h,\mu_s)\, \mathbf{H}^{(0)}(M,\beta_t,\theta,\mu_h)\,\mathbf{u}^{\dagger\, \text{NLL}}(\beta_t,\theta,\mu_h,\mu_s)\,\mathbf{s}^{(0)}\, \bigg\}  \, \nn \\
& \qquad \qquad \quad \times \bigg( \tilde{B}^{(2)}_a(\partial_{\eta_B} + L_B,z_a,\mu_B) \, \tilde{B}^{(0)}_b(z_b,\mu_B) + \tilde{B}^{(0)}_a(z_a,\mu_B) \, \tilde{B}^{(2)}_b(\partial_{\eta_B'} + L_B,z_b,\mu_B)\, \nn \\
& \qquad \qquad \quad + \tilde{B}^{(1)}_a(\partial_{\eta_B} + L_B,z_a,\mu_B) \, \tilde{B}^{(1)}_b(\partial_{\eta_B'} + L_B,z_b,\mu_B)
\bigg) \, \nn \\
& + \frac{\alpha_s(\mu_h) \alpha_s(\mu_B)}{16 \pi^2} \, \mathrm{Tr}\bigg\{\mathbf{u}^{\text{NLL}}(\beta_t,\theta,\mu_h,\mu_s)\, \mathbf{H}^{(1)}(M,\beta_t,\theta,\mu_h)\,\mathbf{u}^{\dagger\, \text{NLL}}(\beta_t,\theta,\mu_h,\mu_s)\,\mathbf{s}^{(0)}\, \bigg\} \, \nn \\
& \qquad  \qquad \quad \times \bigg(\tilde{B}^{(1)}_a(\partial_{\eta_B} + L_B,z_a,\mu_B) \, \tilde{B}^{(0)}_b(z_b,\mu_B) + \tilde{B}^{(0)}_a(z_a,\mu_B) \, \tilde{B}^{(1)}_b(\partial_{\eta_B'} + L_B,z_b,\mu_B)\bigg) \, \nn \\
& + \frac{\alpha_s(\mu_s) \alpha_s(\mu_B)}{16 \pi^2} \,
\mathrm{Tr}\bigg\{\mathbf{u}^{\text{NLL}}(\beta_t,\theta,\mu_h,\mu_s)\, \mathbf{H}^{(0)}(M,\beta_t,\theta,\mu_h)\,\mathbf{u}^{\dagger\, \text{NLL}}(\beta_t,\theta,\mu_h,\mu_s) \,\nn \\
& \qquad \qquad \qquad \qquad \times \,\,\tilde{\mathbf{S}}^{(1)}_{B}(\partial_{\eta_s}+L_s,\beta_t,\theta,\mu_s)\, \bigg\}\, \nn \\
& \qquad \qquad \times \bigg(\tilde{B}^{(1)}_a(\partial_{\eta_B} + L_B,z_a,\mu_B) \, \tilde{B}^{(0)}_b(z_b,\mu_B) + \tilde{B}^{(0)}_a(z_a,\mu_B) \, \tilde{B}^{(1)}_b(\partial_{\eta_B'} + L_B,z_b,\mu_B)\bigg) \, \Bigg\} \, \nn \\
& \qquad \qquad \times \frac{\eta^{\text{NLL}}_{\mathrm{tot}}}{\tau_B^{1-\eta^{\text{NLL}}_{\mathrm{tot}}}}\frac{e^{-\gamma_E\eta^{\text{NLL}}_{\mathrm{tot}}}}{\Gamma( 1 + \eta^{\text{NLL}}_{\mathrm{tot}})}\, .
\end{align}

In the previous equation, we only included the $\ord{\as^2}$ contributions
coming from the boundary conditions of the hard, soft and beam functions
evaluated at their characteristic scales. We neglect other contributions
from {\it e.g.} the evolution matrices, which appear at the same order in the
counting $\as L\sim 1$, since these terms form part of the N$^3$LL resummation.

At present, our implementation of \eq{NNLLcorras2} misses the term in the
second line since we have set the two-loop hard function coefficient
$\mathbf{H}^{(2)}(M,\beta_t,\theta,\mu_h) = 0$. The term in the third line is
only missing the $\tilde{\mathbf{S}}^{(2,0)}_B$ ({\it cfr.} \eq{softcoeffs})
contributions to the two-loop soft functions $\tilde{\mathbf{S}}^{(2)}_{B}(\partial_{\eta_s}+L_s,\beta_t,\theta,\mu_s)$; these are completely unknown at
present and hence have also been set to zero.
We have, however, included the known  two-loop contributions to the beam functions.
As a result, we cannot yet claim full NNLL$^\prime$ accuracy. We define
our results obtained dropping the aforementioned two-loop hard and soft
contributions proportional to $\delta(\Tau_0)$ as an approximate next-to-next-to-leading logarithmic accuracy  (NNLL$^\prime_{\mathrm{a}}$) -- this expression is nevertheless sufficient to match the resummed piece to the NLO fixed order calculations for \ttbarj production.
Indeed, when the evolution is switched off (i.e. set to unity), the formula above reproduces the complete singular spectrum for $\Tau_0 > 0$, valid in the limit $\Tau_0 \ll M$.

\section{Numerical results}
\label{sec:resumresults}
We now present numerical results from the resummation framework developed in
\sec{resummation}. We have implemented \eq{nllformula}, \eq{NNLLcorr} and \eq{NNLLcorras2} in the
\geneva code, as well as the NLO fixed order calculations for \ttbar and
$t\bar{t}$+jet. This allows us to study both the resummed distributions and
the matching of the resummed calculation to the fixed order.
For sake of definiteness, all the results presented in this section
have been obtained for $pp$ collisions at a centre-of-mass energy of
$\sqrt{S} = 13$~TeV and using \texttt{PDF4LHC15\_nnlo} parton
distribution functions from LHAPDF~\cite{Butterworth:2015oua,Buckley:2014ana}. The
central predictions have been obtained running all scales to a common
scale $\mu$ equal to the \ttbar invariant mass $M$.
In all figures
present in this section, the statistical uncertainties associated with
the Monte Carlo integrations are reported, when visible, as vertical
error bars.
We estimate the theoretical uncertainties for the fixed order predictions by
varying the central choice for $\mu_R=\mu_F=M$ up and down by a factor of
two and take the maximal
absolute deviation from the central result as the fixed order uncertainty.

\begin{figure}[tp]
	\begin{center}
		\begin{tabular}{ccc}
			\includegraphics[width=\rescaletwoplots]{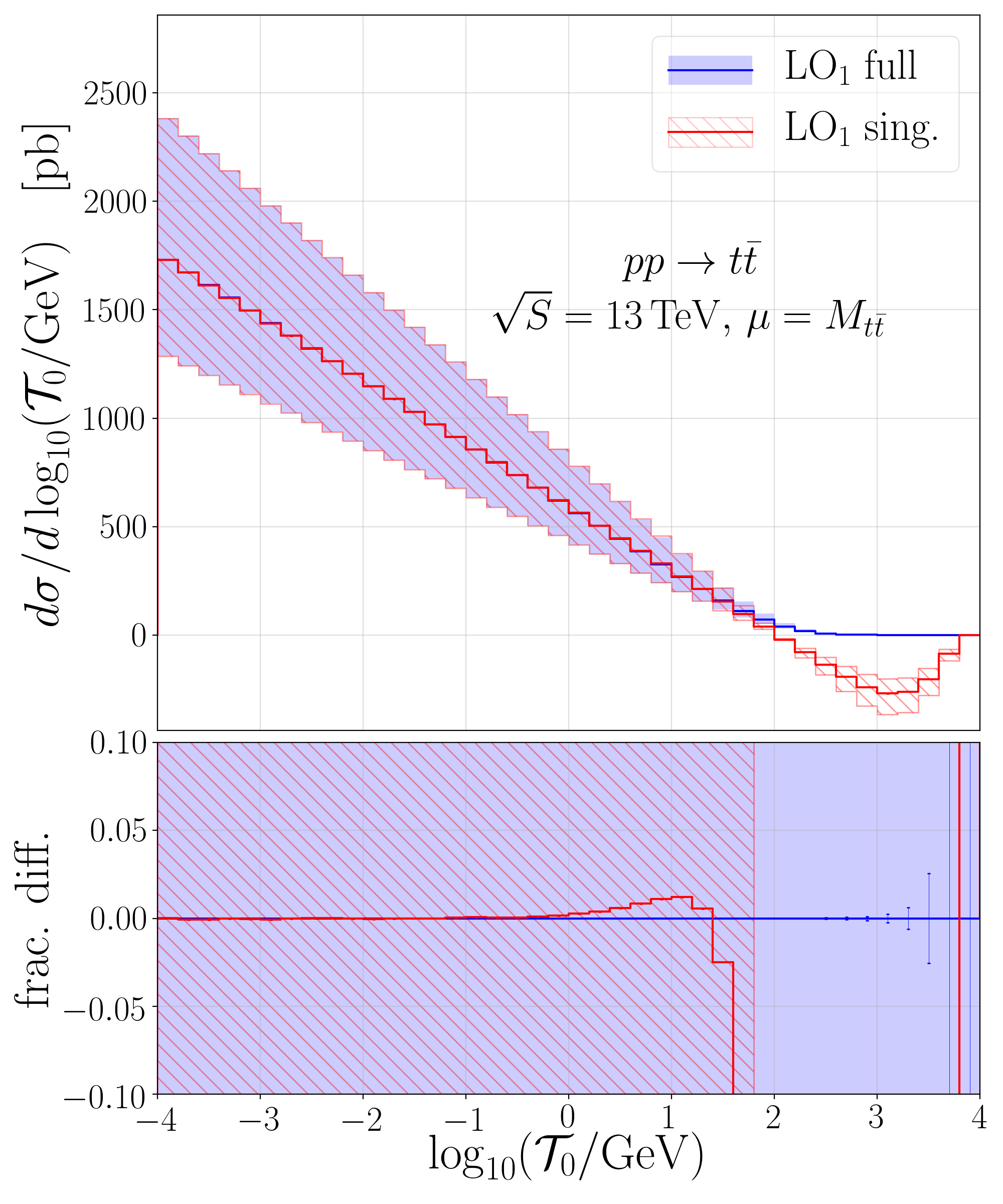} &\hspacebetweentwoplots&
			\includegraphics[width=\rescaletwoplots]{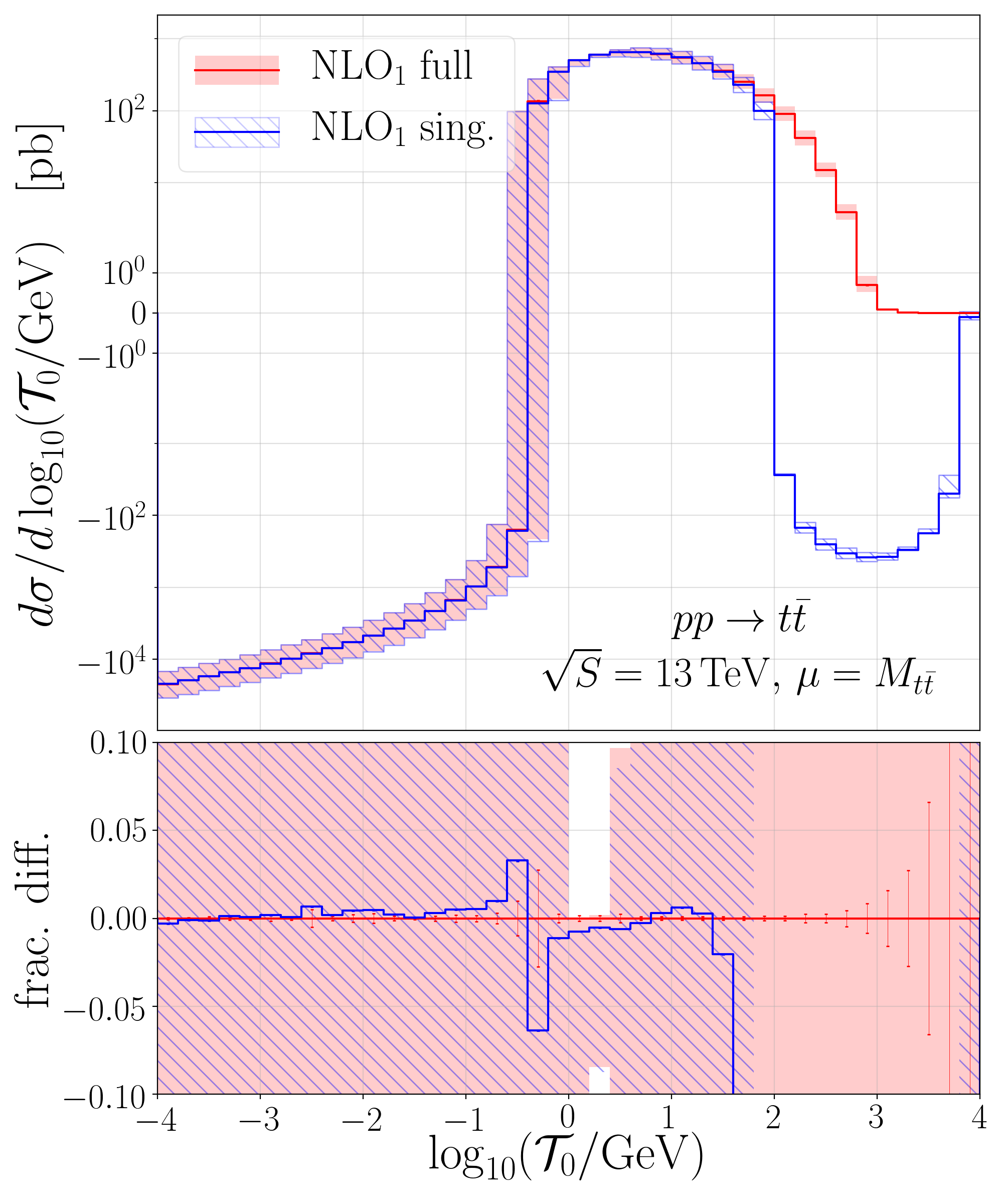}
		\end{tabular}
	\end{center}
	\spaceabovefigurecaption
	\caption{
          Approximate fixed order results for the $\Tau_0$ distribution obtained from our factorisation theorem compared with full calculations at LO (left) and NLO (right). The approximate results correctly reproduce the fixed order behaviour in the $\Tau_0 \to 0$ limit.
		\label{fig:log10Tau0SingvsFO}
	}
	\spacebelowfigurecaption
\end{figure}

We begin by verifying that the approximate fixed order expressions, which we
obtain from the resummed calculation by setting the various resummation scales
equal to the hard scale, are able to
reproduce the behaviour of the full fixed order calculation as $\Tau_0\to
0$. Comparisons of the full with the approximate fixed order results are shown
in \fig{log10Tau0SingvsFO} at LO$_1$ (i.e. LO $t\bar{t}$+jet) and NLO$_1$
accuracy. We observe that, for small values of $\Tau_0 \lesssim 10^{-1}$ GeV,
the approximate FO reproduces the behaviour of the full calculation very well, both for the central values and the scale variations.
This gives us confidence that the factorisation theorem is valid and that our calculation of the finite part of the one-loop soft function is correct.
We notice that when the full NLO$_1$ result crosses zero in the right plot, the associated statistical errors grow large, resulting in a instability in the ratio plot shown in the lower panel.

\begin{figure}[tp]
	\begin{center}
		\begin{tabular}{ccc}
			\includegraphics[width=\rescaletwoplots]{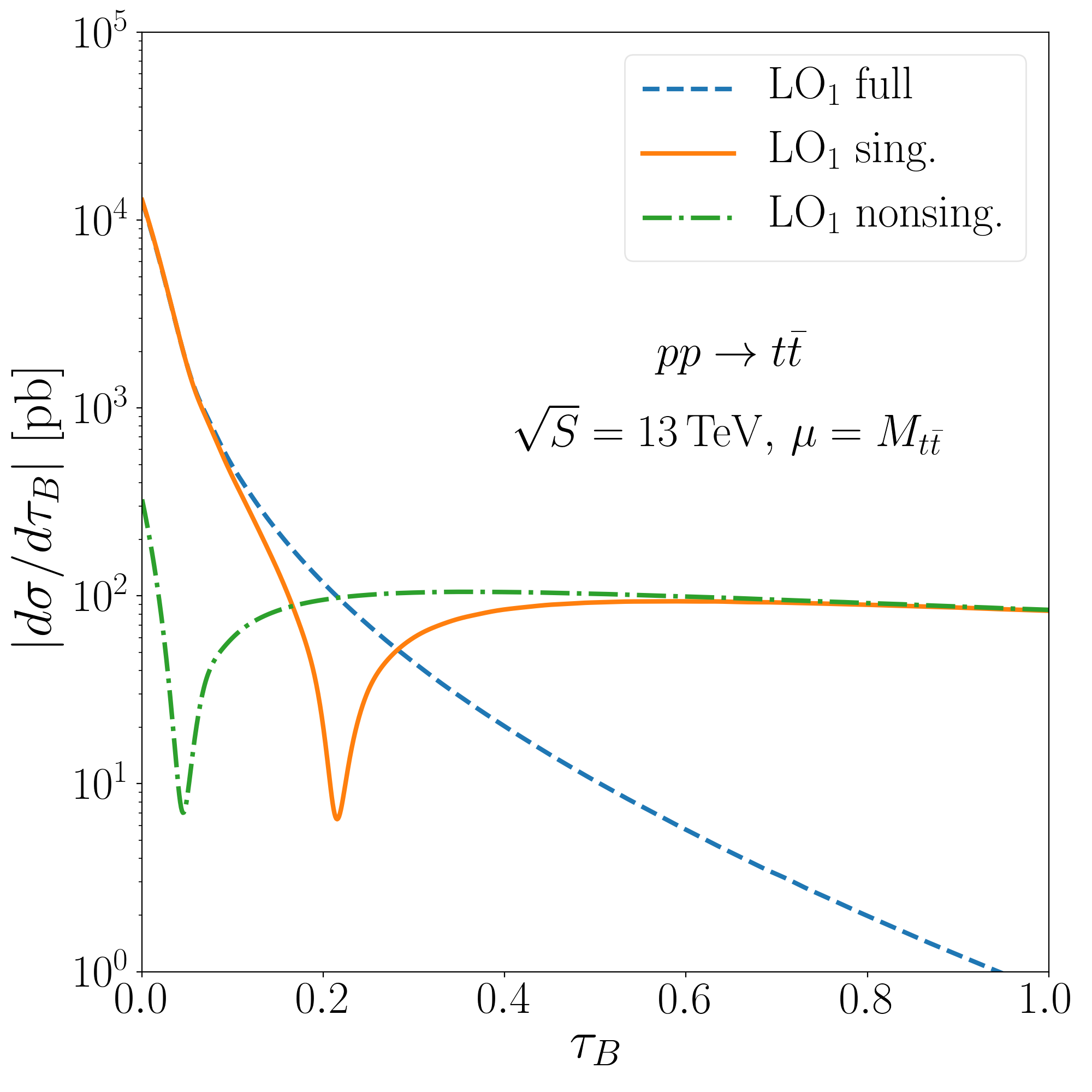} &\hspacebetweentwoplots&
			\includegraphics[width=\rescaletwoplots]{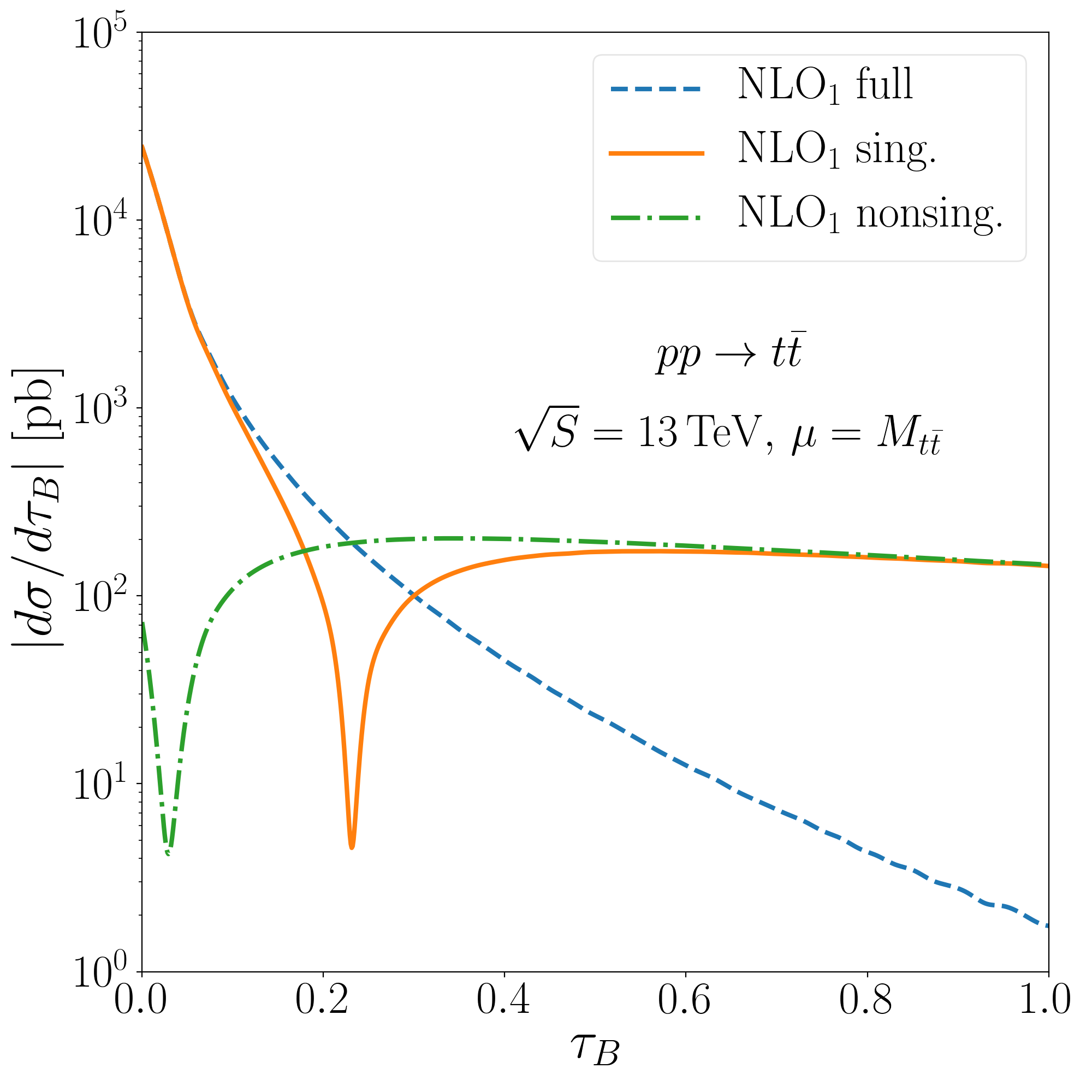}
		\end{tabular}
	\end{center}
	\spaceabovefigurecaption
	\caption{
          Comparison of the absolute values for the singular and nonsingular contributions to the $\Tau_0$ distribution with fixed order results at LO (left) and NLO (right) accuracy.
		\label{fig:tau0SingvsNonSingvsFO}
	}
	\spacebelowfigurecaption
\end{figure}

Before studying the resummed result, we have to provide a procedure to turn off the
resummation before the exponentiated singular terms become too large, spoiling the predictions
in the fixed order region. We do so in a smooth fashion by employing the
profile scales introduced in Refs.~\cite{Ligeti:2008ac,Abbate:2010xh,Berger:2010xi}. These profiles evolve the beam and soft scales to the hard scale as a function of $\tau_B$ and hence stop the RG evolution and resummation when the common scale $\mu_{\rm{NS}} = \mu_S = \mu_B = \mu_H$ is reached. Specifically, the profiles take the form:
\begin{align} \label{eq:centralscale}
\mu_H &= \mu_{\rm{NS}}\nn\,,  \\
\mu_S(\Tau_0) & = \mu_\NS\ f_{\rm run}(\Tau_0/M)\,, \\
\mu_B(\Tau_0) &=  \mu_\NS\ \sqrt{f_{\rm run}(\Tau_0/M)} \,,\nn
\end{align}
where the common profile function $f_{\rm run}(y)$ is given by~\cite{Stewart:2013faa}
\begin{align}
f_{\rm run}(y) &=
\begin{cases} y_0 \bigl[1+ (y/y_0)^2/4 \bigr] & y \le 2y_0\,,
\\ y & 2y_0 \le y \le y_1\,,
\\ y + \frac{(2-y_2-y_3)(y-y_1)^2}{2(y_2-y_1)(y_3-y_1)} & y_1 \le y \le y_2\,,
\\  1 - \frac{(2-y_1-y_2)(y-y_3)^2}{2(y_3-y_1)(y_3-y_2)} & y_2 \le y \le y_3\,,
\\ 1 & y_3 \le y\,.
\end{cases}
\label{eq:frun}
\end{align}
This functional form ensures the canonical scaling behaviour  for values below $y_1$ and turns off the resummation above $y_3$. In order to determine the parameters $y_i$ of the profiles, it is instructive to examine the behaviour of the singular and nonsingular contributions to the cross section as a function of $\tau_B$ relative to the fixed order calculation. This is shown at LO$_1$ and NLO$_1$ accuracy in \fig{tau0SingvsNonSingvsFO}. We see that the singular contribution to the cross section becomes of a similar size to the fixed order when $\tau_B$ is just above 0.2. The behaviour at different orders is very similar. We therefore make the choices
\begin{align} \label{eq:TauBprofile}
y_0 = 1.0\GeV/M\,, \quad
\{y_1,y_2, y_3\} = \{0.1, 0.175, 0.25\}\,.
\end{align}

We now discuss the resummed results. In order to estimate the theoretical uncertainties, we
vary the central choices for the 
profile scales in \eq{centralscale} independently while keeping
the hard scale fixed. This gives us four independent variations. In
addition, we consider two more profile functions where we shift all
the $y_i$ transition points together by $\pm 0.05$ while keeping all
of the scales fixed at their central values. Hence, we
obtain in total six profile variations. We consider the maximal
absolute deviation in the results with respect to the central
prediction as the resummation uncertainty. 

\begin{figure}[tp]
	\begin{center}
		\begin{tabular}{ccc}
			\includegraphics[width=\rescaletwoplots]{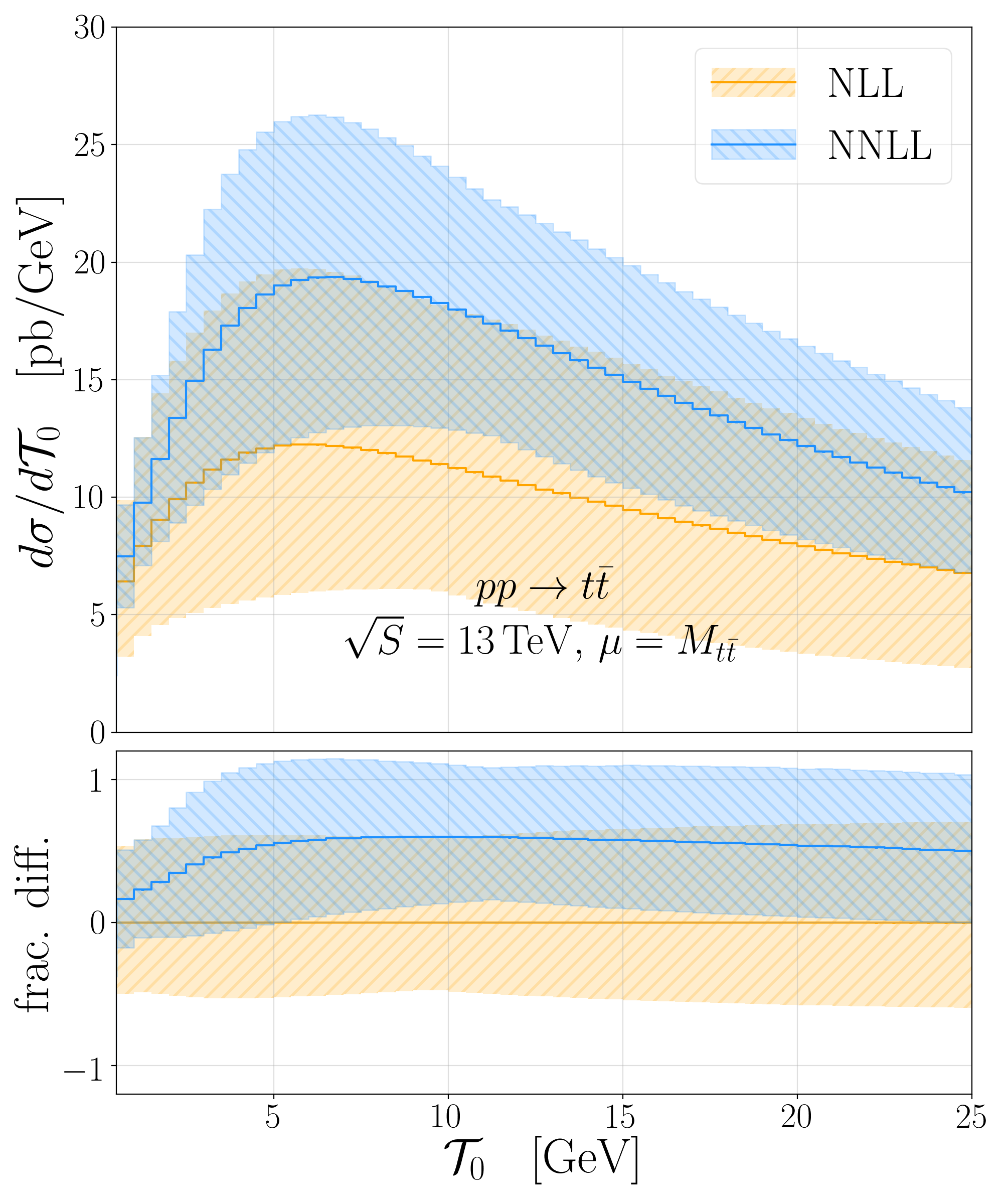} &\hspacebetweentwoplots&
			\includegraphics[width=\rescaletwoplots]{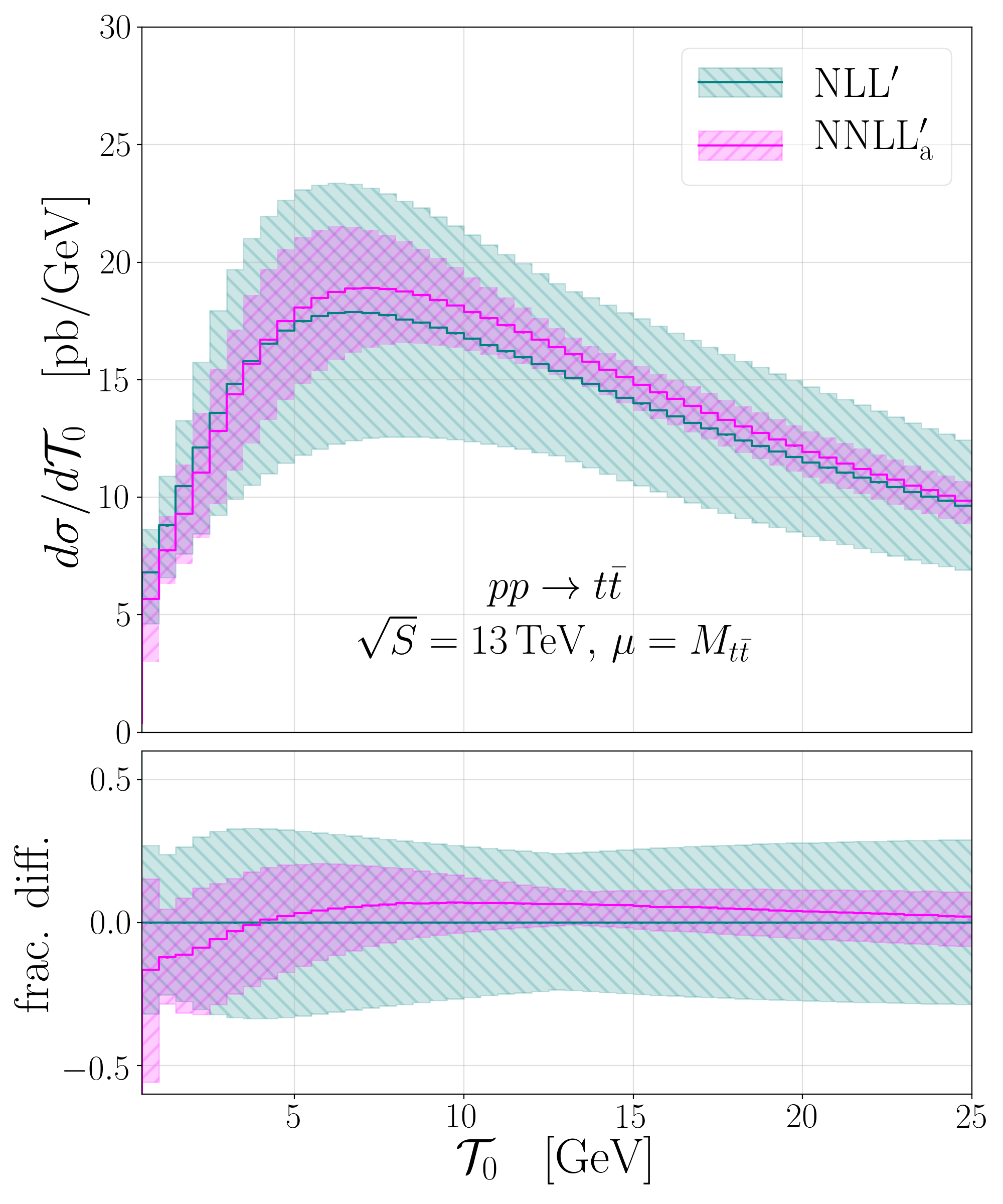}
		\end{tabular}
	\end{center}
	\spaceabovefigurecaption
	\caption{Resummed $\Tau_0$ distribution at successive unprimed (left) and primed (right) orders. Compared to the full NNLL$^\prime$ result, the approximate NNLL$^\prime_{\rm{a}}$ prediction shown on the right misses only  finite  $\ord{\as^2}$ terms proportional to $\delta(\Tau_0)$ in the hard and soft functions.
		\label{fig:resummedconvplots}
	}
	\spacebelowfigurecaption
\end{figure}

In \fig{resummedconvplots}, we show the peak region of the resummed $\Tau_0$
distribution.
We compare predictions at different primed and unprimed levels
of accuracy from NLL to NNLL$^\prime_{\rm{a}}$.
Examining the unprimed
results, we see a large shift in the central value between the NLL and NNLL
results, though the central prediction for the NNLL result remains within the
scale uncertainty band of lower order calculation. We also observe
that the size of the band does not reduce substantially when moving from one order to the 
next. On the other hand, comparing the NLL$^\prime$ and NNLL$^\prime_{\rm{a}}$
results we observe both a more stable central value and also a sizeable
reduction of the theoretical uncertainties. This highlights the need for full
NNLL$^\prime$ accuracy in this process, which we hope to report on in future
work.

\begin{figure}[tp]
	\begin{center}
		\begin{tabular}{ccc}
			\includegraphics[width=\rescaletwoplots]{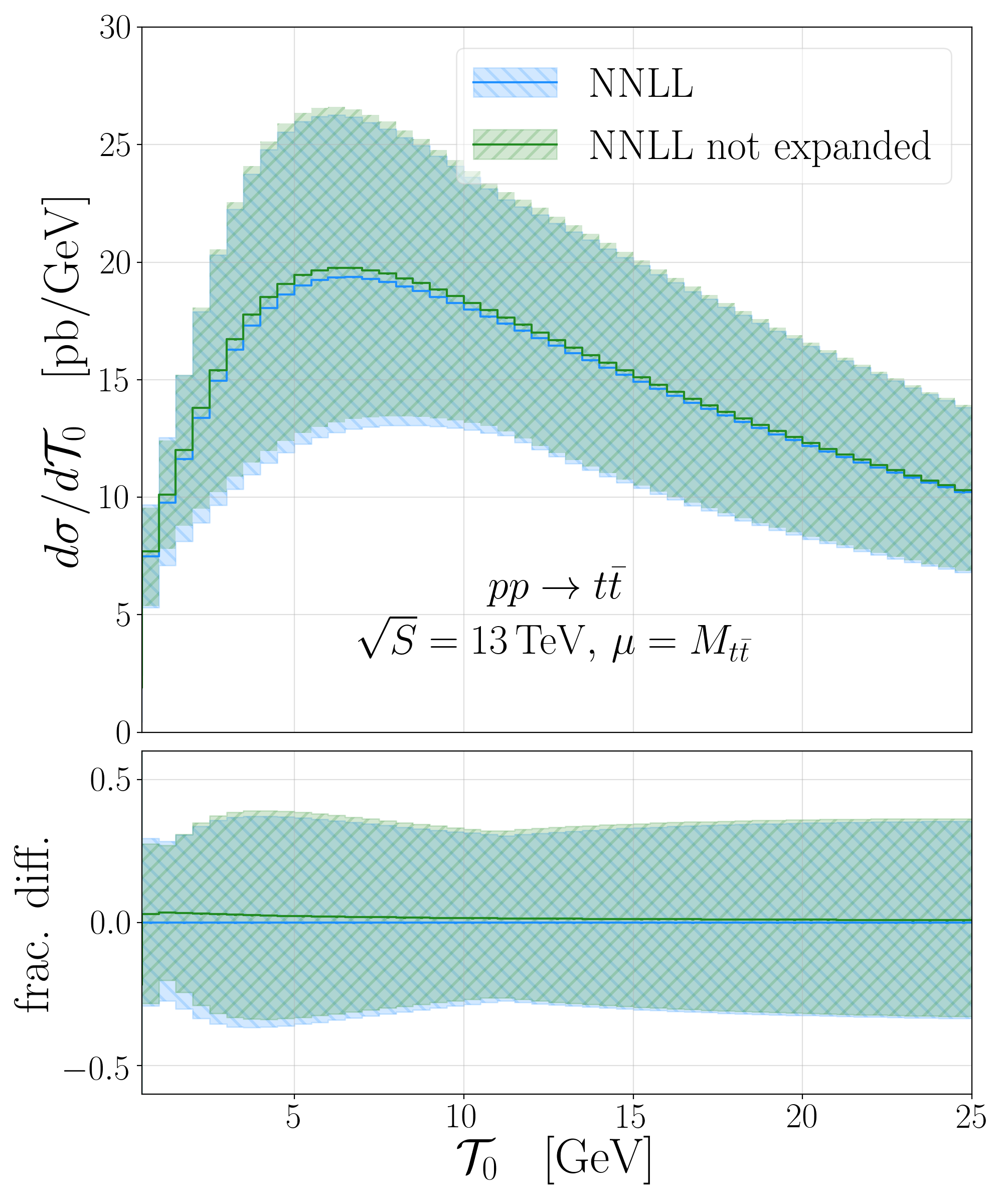} &\hspacebetweentwoplots&
			\includegraphics[width=\rescaletwoplots]{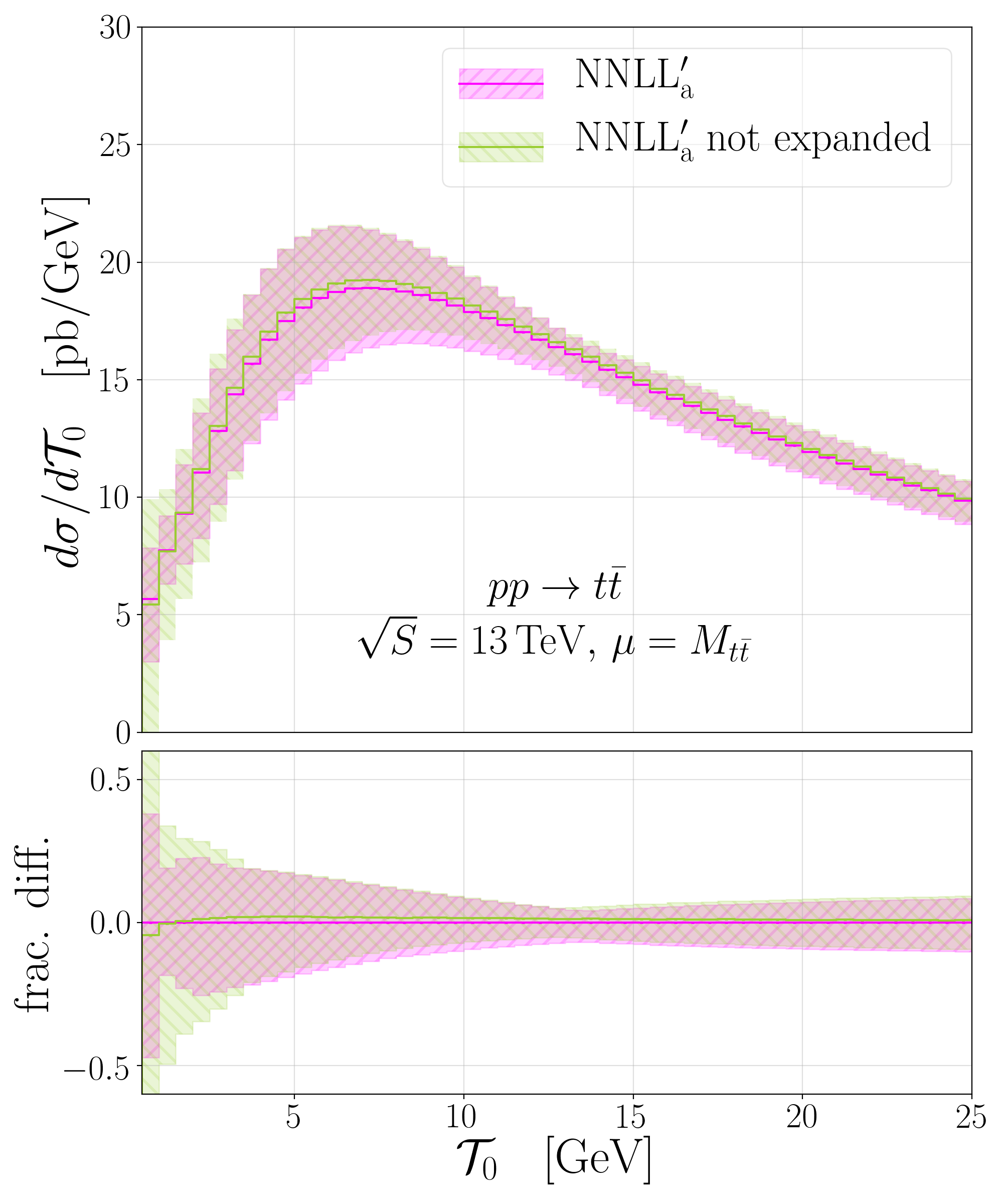}
		\end{tabular}
	\end{center}
	\spaceabovefigurecaption
	\caption{
          Resummed $\Tau_0$ distribution with and without the expansion of $U$ in \eq{diagonalU}, at both NNLL (left) and NNLL$_{\rm{a}}^\prime$ accuracy (right).
		\label{fig:resummedExpvsNotExp}
	}
	\spacebelowfigurecaption
\end{figure}

As mentioned in \sec{nllandnnll}, for the production of coloured
particles there is a certain amount of ambiguity in whether one should
expand terms or instead keep them inside the exponential
prefactor. This ambiguity starts at NNLL accuracy, since these terms
 are the first to contribute at $\ord{\as}$ in the logarithmic counting of 
the exponent.  Indeed,
while it is necessary to evaluate the non-diagonal evolution matrix
$\mathbf{u}$ as a perturbative expansion, the product between the
diagonal evolution matrix $U$ and the generating function appearing
{\it e.g. }in the first line of \eq{NNLLcorr} may be expanded in the
same way or kept exact. We choose the former by default; however, it
is interesting to assess the (formally higher order) effect of making
the other choice. In \fig{resummedExpvsNotExp}, we compare the
resummed distribution with and without this expansion, at both NNLL
and NNLL$_{\rm{a}}^\prime$ accuracy. We observe very little difference
between the expanded and unexpanded results, suggesting that the
effects of these missing higher order terms in the expanded results
are minimal.

We now consider the matching of the resummed and fixed order calculations. We
perform an additive matching, following the same spirit as recent \geneva
implementations (see \textit{e.g.} Ref.~\cite{Alioli:2019qzz}).
The appropriate combinations of resummed and fixed order accuracies are given in \tab{resummedacc}.
 The total perturbative
uncertainty is calculated by adding in quadrature the  previously discussed fixed order and resummation
uncertainties. 
We define our matched
spectrum as
\begin{align}
  \frac{\de\sigma^{\rm match}}{\de \Tau_0} = \frac{\de\sigma^{\rm resum}}{\de \Tau_0} + \frac{\de\sigma^{\rm FO}}{\de \Tau_0} - \left[\frac{\de\sigma^{\rm resum}}{\de \Tau_0}\right]_{\rm FO}\,,
\end{align}
where the final term removes double-counting between the resummed and fixed order pieces.
In \geneva implementations at NNLL$^\prime$+NNLO, it acts as a subtraction term local in $\Tau_0$, which requires the fixed order calculation to use a $\Tau_0$-preserving mapping. This can have the positive feature of reducing the impact of fiducial power corrections compared to a simple slicing approach~\cite{Ebert:2019zkb,Ebert:2020dfc}.

\begin{figure}[tp]
	\begin{center}
		\begin{tabular}{ccc}
			\includegraphics[width=\rescalethreeplots]{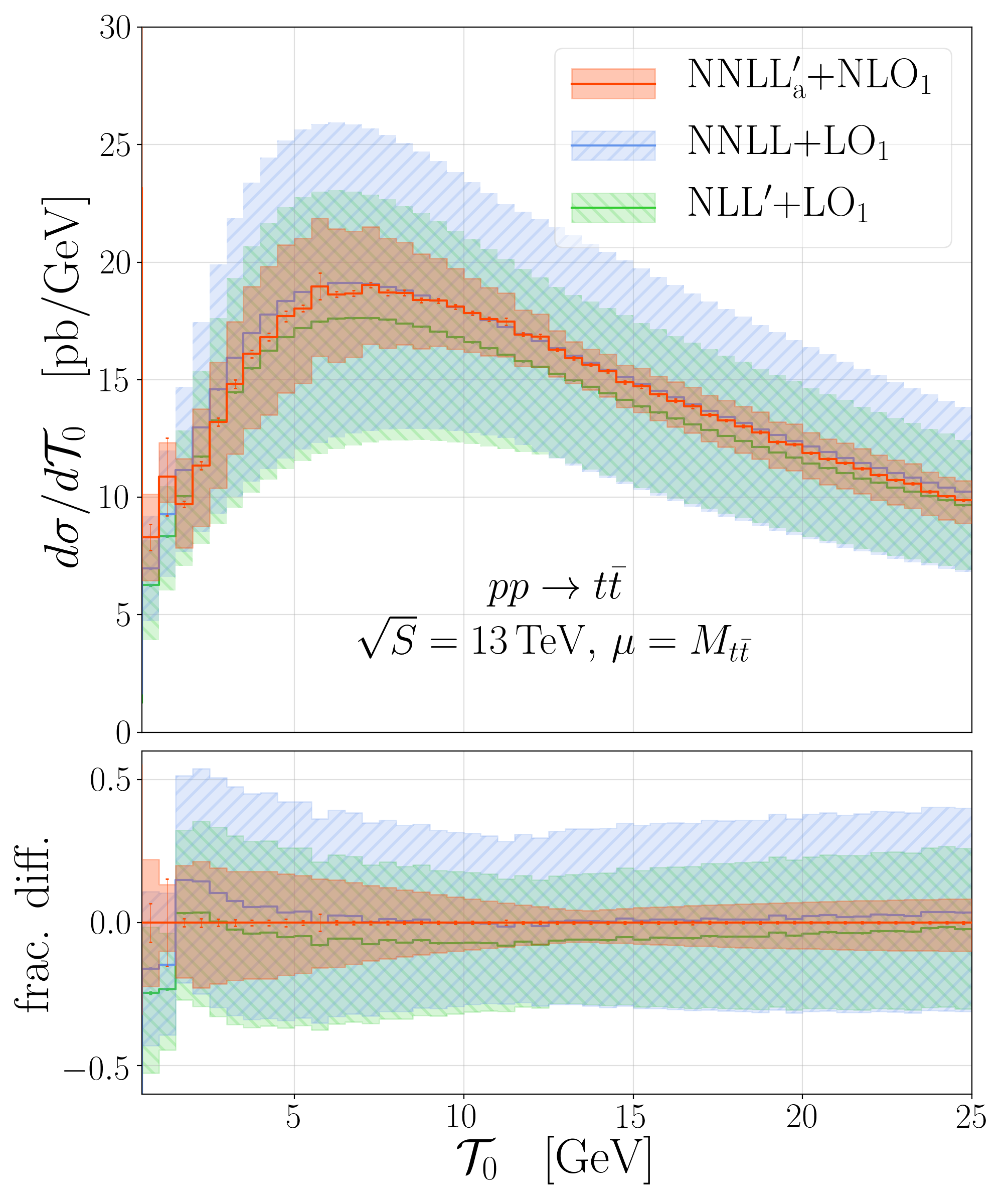} & \includegraphics[width=\rescalethreeplots]{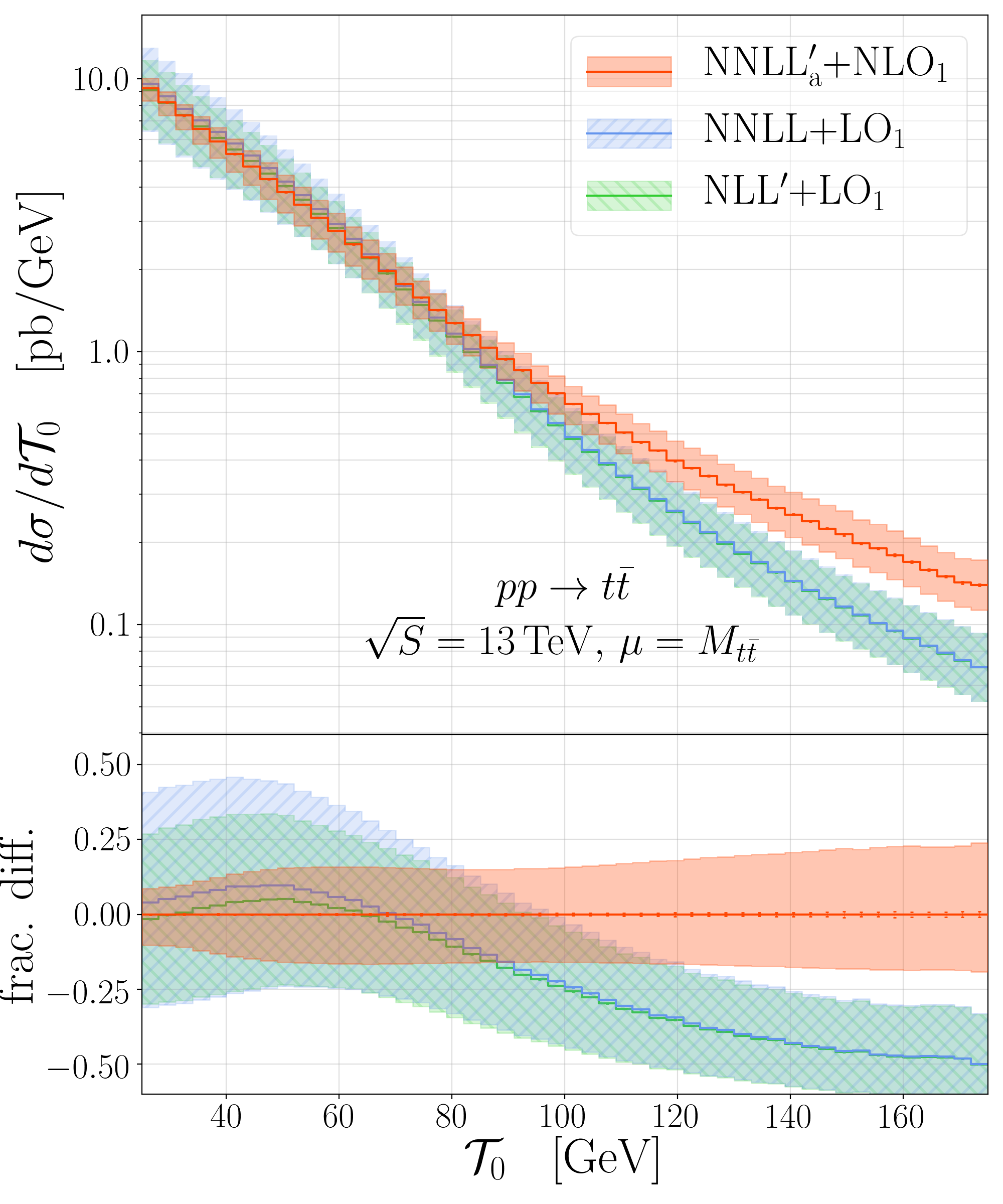} &
			\includegraphics[width=\rescalethreeplots]{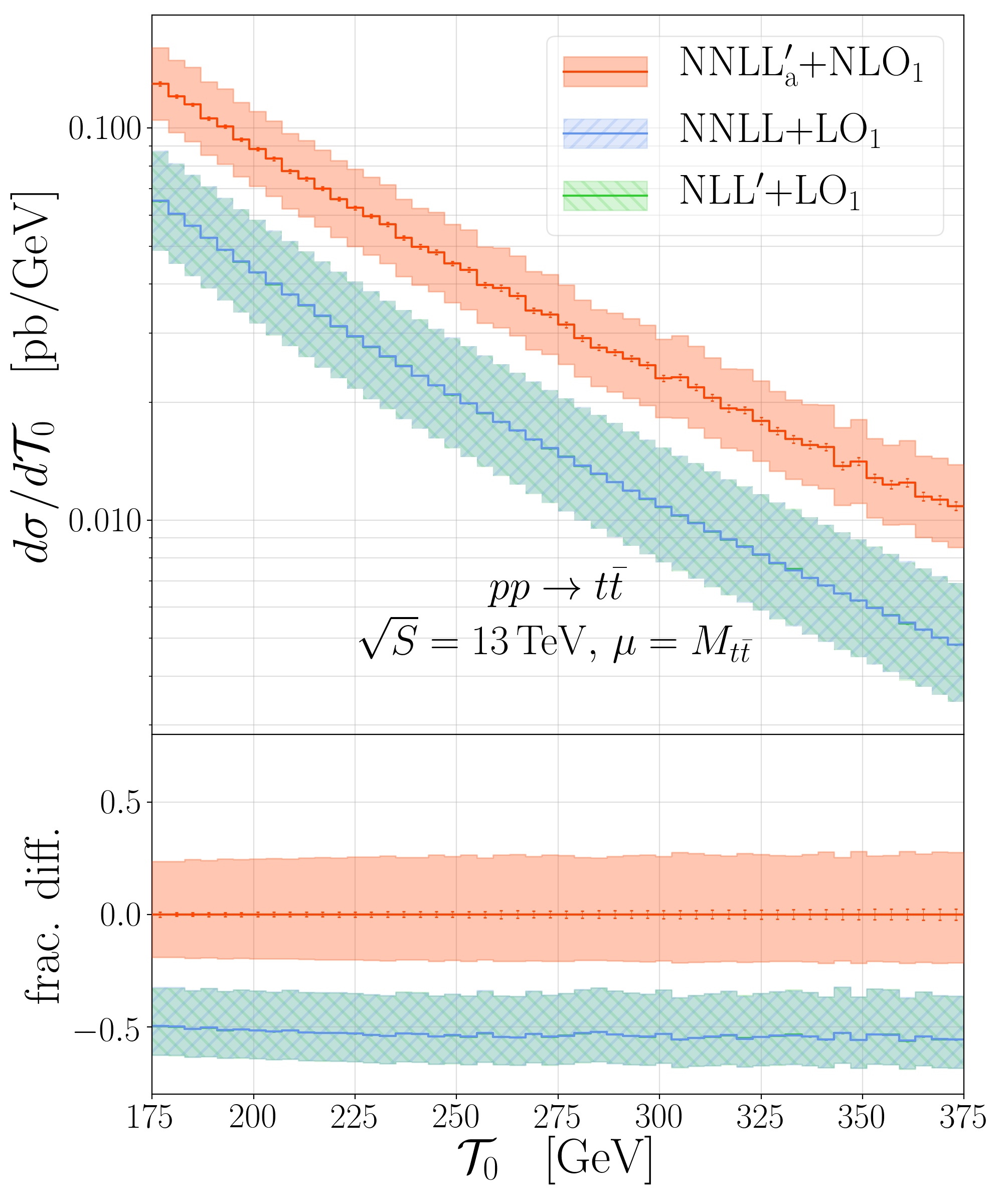}
		\end{tabular}
	\end{center}
	\spaceabovefigurecaption
	\caption{
        Resummed predictions matched to the appropiate fixed order for the $\Tau_0$ distribution  at  increasing accuracy in the peak (left), transition (centre) and tail (right) regions.
		\label{fig:Tau0matched}
	}
	\spacebelowfigurecaption
\end{figure}

Finally, in \fig{Tau0matched} we  present our best predictions across the
whole spectrum. In order to highlight the effect of these higher-order
corrections we show the resummed results at various resummation orders matched
to the appropriate fixed order calculations. We divide the spectrum into the
peak region, where resummation effects are most important, the transition,
where resummed and fixed order contributions compete for importance, and the
tail, where the fixed order is dominant. Examining the peak region, we notice
slightly larger uncertainty bands for the NNLL+LO$_1$ compared to the
NLL$^\prime$+LO$_1$. The uncertainty bands are, however, significantly reduced once
NNLL$_{\rm{a}}^\prime$+NLO$_1$ accuracy is reached. In the transition and tail
regions, a clear difference between the NNLL$_{\rm{a}}^\prime$+NLO$_1$ and the
lower order results emerges above $\sim 60$ GeV due to the additional
contributions of the NLO$_1$ calculation.

\section{Conclusions}
\label{sec:conc}
In this work, we studied the resummation of the zero-jettiness variable $\Tau_0$ in the context of top-quark pair production at hadron colliders.
Starting from the frameworks of the Soft-Collinear and Heavy-Quark effective theories, we derived a factorisation formula for the process, which allowed us to express
the resummed cross section at any logarithmic order in terms of hard, soft and beam functions. We then proceeded to calculate the relevant soft functions at one-loop order, which is a necessary boundary condition for reaching NLL$^\prime$ accuracy. 
Taking advantage of the known behaviour of the large logarithms for each of the ingredients of the factorisation theorem under renormalisation group evolution, we derived final expressions for the $\Tau_0$ distribution at NLL$^\prime$, NNLL and approximate NNLL$^\prime$ accuracy, which we referred to as NNLL$_{\rm{a}}^\prime$.
The approximation neglects the two-loop contributions to the hard and soft functions proportional to $\delta(\Tau_0)$, where the latter are presently unknown.
We numerically evaluated our formul\ae~up to approximate NNLL$^\prime$ order, and matched our resummed predictions to fixed order calculations of the \ttbar and \ttbarj processes at NLO.

The work presented here achieves the highest accuracy matched calculation in a resolution variable for the \ttbar process, and provides the foundation for a full NNLOPS implementation in the \geneva framework. In the future, resummed calculations for other resolution variables (such as the transverse momentum of the \ttbar pair) at this accuracy could also be used in \geneva to examine the effect of the choice of resolution variable on NNLOPS predictions for this process, as was done for Drell-Yan in Ref.~\cite{Alioli:2021qbf}. 

At present, two ingredients are missing from our implementation which prevent us from reaching full NNLL$^\prime$ accuracy -- the colour-decomposed hard function at NNLO, and the term of the two-loop soft function proportional to $\delta(\Tau_0)$. The first of these can, in principle, be constructed from ingredients already available in the literature -- for example, the necessary two-loop amplitudes in colour-decomposed form were published in Ref.~\cite{Chen:2017jvi} in the form of interpolation tables. The second, soft piece, could either be calculated analytically or be obtained using automated numerical tools such as \texttt{SoftSERVE}~\cite{Bell:2018vaa,Bell:2018oqa,Bell:2020yzz}. We intend to explore the effect of incorporating these additional terms on our results in the near future.

The factorisation theorem which we derived in this work is also applicable to other processes which involve the production of heavy coloured particles, such as the associated production of a heavy quark pair with Higgs or electroweak bosons.
The main obstacle in reaching NNLL$^\prime$ accuracy for these processes is the availability of two-loop hard and soft functions. Nevertheless, in order to achieve the approximate NNLL$^\prime_{\mathrm {a}}$ accuracy used for our results, one can instead make use of the one-loop hard functions which are available in the literature~\cite{Broggio:2015lya,Broggio:2016zgg,Broggio:2016lfj,Broggio:2017kzi}. The one loop soft functions must be calculated anew, allowing for a more general kinematic dependence.

\section*{Acknowledgements}
\label{sec:Acknowledgements}

We are grateful to S.~Kallweit, R.~Nagar and D.~Napoletano for 
useful discussions. We thank A.~Gavardi, G.~Marinelli and F.~Tackmann for a careful reading of and comments on the manuscript.
The work of SA, AB and MAL is supported by the ERC Starting Grant REINVENT-714788.
SA acknowledges funding from Fondazione Cariplo and Regione
Lombardia, grant 2017-2070 and from MIUR
through the FARE grant R18ZRBEAFC. MAL is also supported by the Deutsche
Forschungsgemeinschaft (DFG) under Germany's Excellence Strategy -- EXC 2121 ``Quantum Universe''
 -- 390833306. We acknowledge the CINECA award
under the ISCRA initiative and the National Energy Research Scientific
Computing Center (NERSC), a U.S. Department of Energy Office of
Science User Facility operated under Contract No. DEAC02-05CH11231,
for the availability of the high performance computing resources
needed for this work.

\appendix

\section{Derivation of the factorisation formula}
\label{app:Hvqfactheorem}

In this appendix we discuss the details of the derivation of the factorisation formula for the top-quark pair production process in the limit of small $\Tau_0$ by employing position space SCET and the multipole expansion.

We begin with some definitions which will be needed throughout this appendix. We work in the centre-of-mass frame of the hadronic collisions  and introduce two light-like vectors $n_a$ and $n_b$ along the beam directions so that the momenta of the incoming protons are given by $P^\mu_{a,b}=E_{\mathrm{cm}}\, n^\mu_{a,b}/2$. These vectors satisfy the relations $n_b = \bar{n}_a=(1,-\vec{n}_a)$ and $n_a\cdot n_b=2$.

Neglecting all complications related to Glauber modes, the description of the top-quark pair production process within an effective field theory framework requires both SCET, to describe the interactions of the collinear fields with the soft gluons, and Heavy-Quark Effective Theory (HQET), to describe the soft interactions of the heavy quark fields. When the heavy-quark fields $h_v(x)$ appear in combination with collinear (anti-collinear) fields, they must be Taylor expanded around $x^-$ ($x^+$) since the residual $x$ variations of these fields are identical to those of ultrasoft fields. The effective theory operators which are relevant for this process contain two heavy quark fields, one collinear and one anti-collinear field: the double projection onto plus and minus components therefore implies that the heavy quark fields must be evaluated at the origin.
While the final state heavy quarks by definition do not contribute to the radiation in any hemisphere, the soft Wilson lines along the directions $v_3$ and $v_4$, which arise from the decoupling of the soft gluon interactions from the heavy quark fields, contribute instead to the hemisphere radiation measurements.
The effective Hamiltonian for this process reads \cite{Ahrens:2010zv}
\begin{align}
\mathcal{H}_{\rm eff}(x)=\sum_{I,m}\int \de r \,\de t \,\,e^{im_t(v_3+v_4)\cdot x} \big[ C^{gg}_{Im}(r,t)\,O^{gg}_{Im}(x,r,t) + C^{q\bar{q}}_{Im}(r,t)\,O^{q\bar{q}}_{Im}(x,r,t) + (q\leftrightarrow \bar q) \big]\,,
\end{align}
where $I$ labels the colour and $m$ the Dirac structures. The operators explicitly read
\begin{align}\label{eq:operators}
O^{q\bar{q}}_{Im}(x,r,t)&=\sum_{\{a\}}(c_I^{q\bar{q}})_{\{a\}}\, \bar{\chi}^{a_2}_{\bar{c}}(x+rn_a)\,\Gamma^\prime_m\, \chi^{a_1}_{c}(x+tn_b)\,\bar{h}^{a_3}_{v_3}(x)\,\Gamma^{\prime\prime}_m\, h^{a_4}_{v_4}(x) \nn \\
O^{gg}_{Im}(x,r,t)&=\sum_{\{a\}}(c_I^{gg})_{\{a\}}\, \mathcal{A}^{a_2}_{\bar{c}\nu\perp}(x+rn_a) \mathcal{A}^{a_1}_{c\mu\perp}(x+tn_b)\,\bar{h}^{a_3}_{v_3}(x)\,\Gamma^{\mu\nu}_m \,h^{a_4}_{v_4}(x)\,,
\end{align}
where the fields $\chi_{c}$ and $\mathcal{A}_{c}$ are the collinear gauge-invariant, collinear building blocks
\begin{align}
\chi_{c} = W^\dagger_{c}(x) \xi_{c}(x), \qquad \mathcal{A}^\mu_{c\, \perp} = W^\dagger_{c}(x)\big(i D^\mu_\perp\, W_{c}(x)\big)\,,
\end{align}
with $\xi_{c}= \big(\slashed{n}_a \slashed{\bar{n}}_a/4\big) \, \psi(x)$. Similar equations hold for the anti-collinear quark $\chi_{\bar{c}}$ and gluon $\mathcal{A}^\mu_{\bar{c}}$ building blocks.
The structures $\Gamma^{\mu \nu}_m$, $\Gamma^\prime_m$ and $\Gamma^{\prime \prime}_m$ span all relevant combinations of Dirac matrices and external vectors.
The $c_I$ tensors define the orthogonal colour basis
\begin{align}\label{eq:colorbasis}
(c_1^{q\bar{q}})_{\{a\}}=\delta_{a_1a_2}\delta_{a_3a_4}\,,\quad & (c_2^{q\bar{q}})_{\{a\}}=t^c_{a_2a_1}t^c_{a_3a_4}\,,\nn\\
(c_1^{gg})_{\{a\}}=\delta^{a_1a_2}\delta_{a_3a_4}\,,\quad (c_2^{gg})_{\{a\}}=i&f^{a_1a_2c}t^c_{a_3a_4}\,,\quad (c_3^{gg})_{\{a\}}=d^{a_1a_2c}t^c_{a_3a_4}\,
\end{align}
where ${\{a\}}=\{a_1,a_2,a_3,a_4\}$ is a set of colour indices which can be either in the fundamental or adjoint representation depending on the field with which they are contracted.
It is convenient to decouple the soft gluon interactions via the BPS (decoupling) transformations ~\cite{Bauer:2001yt} for the collinear and heavy quark fields and obtain for the $q\bar{q}$ channel
\begin{align}
O^{q \bar{q}}_{I\, m}(x,r,t) &= \sum_{\{a\},\{b\}}\,\big(c^{q\bar{q}}_I\big)_{\{a\}}\,\big[O^h_m(x)\big]^{b_3 b_4}\,\big[O^c_m(x,r,t)\big]^{b_1 b_2}\,\big[O^s(x)\big]^{\{a\},\{b\}}\, ,
\end{align}
where
\begin{align}
\big[O^h_m(x)\big]^{b_3 b_4} &= \bar{h}^{b_3}_{v_3}(x)\,\Gamma^{\prime\prime}_m\, h^{b_4}_{v_4}(x) ,\quad \quad \big[O^c_m(x,r,t)\big]^{b_1 b_2} = \bar{\chi}^{b_2}_{\bar{c}}(x+rn_a)\,\Gamma^\prime_m\, \chi^{b_1}_{c}(x+tn_b) \, , \nonumber \\
&\big[O^s(x)\big]^{\{a\},\{b\}} = \big[Y^\dagger_{v_3}(x)\big]^{b_3 a_3}\big[Y_{v_4}(x)\big]^{a_4 b_4}\big[Y^\dagger_{n_b}(x)\big]^{b_2 a_2}\big[Y_{n_a}(x)\big]^{a_1 b_1}\, ,
\end{align}
and the soft Wilson lines are defined as
\begin{align}
\big[Y_{n_{\{a,b\}}}(x)\big]^{cd}& = \mathcal{P} \, \mathrm{exp} \bigg( i g_s \int^0_{-\infty}\, dt \, n_{\{a,b\}}\cdot A^e_s(x+t n_{\{a,b\}})\, t^e_{cd}\bigg) \, , \\
\big[Y_{v_{\{3,4\}}}(x)\big]^{cd} & = \mathcal{P}\, \mathrm{exp} \bigg( - i g_s \int^\infty_{0}\, dt \, v_{\{3,4\}}\cdot A^e_s(x+t v_{\{3,4\}})\, t^e_{cd}\bigg)\, .
\end{align}
For the gluon fusion channel we have instead
\begin{align}
O^{gg}_{I\, m}(x,r,t) & = \sum_{\{a\},\{b\}}  \big(c^{gg}_I\big)_{\{a\}}\, \big[O^h_m(x)\big]^{\mu \nu}_{b_3 b_4} \big[O^c(x,r,t)\big]^{b_1 b_2}_{\mu\nu} \big[O^s(x)\big]^{\{a\},\{b\}}\, ,
\end{align}
where
\begin{align}
\big[O^h_m(x)\big]^{\mu \nu}_{b_3 b_4}& = \bar{h}^{b_3}_{v_3}(x)\,\Gamma^{\mu\nu}_m \,h^{b_4}_{v_4}(x),\quad \quad \big[O^c(x,r,t)\big]^{b_1 b_2}_{\mu\nu} = \mathcal{A}^{b_1}_{c\mu\perp}(x+tn_b) \mathcal{A}^{b_2}_{\bar{c}\nu\perp}(x+rn_a)\, \nonumber \\
& \big[O^s(x)\big]^{\{a\},\{b\}} =  \big[Y^\dagger_{v_3}(x)\big]^{b_3 a_3}\big[Y_{v_4}(x)\big]^{a_4 b_4}\big[Y^{\mathrm{adj}\,\dagger}_{n_b}(x)\big]^{b_2 a_2}\big[Y^{\mathrm{adj}}_{n_a}(x)\big]^{a_1 b_1} \, ,
\end{align}
and we have used the ``adj" superscript for the Wilson line in the adjoint representation
\begin{align}
\big[Y^{\mathrm{adj}}_{n_{\{a,b\}}}(x)\big]^{cd}& = \mathcal{P} \, \mathrm{exp} \bigg( i g_s \int^0_{-\infty}\, dt \, n_{\{a,b\}}\cdot A^e_s(x+t n_{\{a,b\}})\, \big(-i f^{ecd}\big)\bigg) \, .
\end{align} 

We consider $B^\mu_a$ and $B^\mu_b$ to be the total final-state hadronic
momenta in the hemispheres $a$ and $b$ and introduce their two components
$B_a^+=n_a\cdot B_a$, $B_b^+=n_b\cdot B_b$. We evaluate the cross
section differential in $B_a^+$ and $B_b^+$ in the limit where the radiation
is soft or collinear to the initial beam directions and introduce a sum $\displaystyle \sum_X\!\!\!\!\!\!\!\!\!\int $ over all possible radiation states

\begin{align}\label{eq:crosssection}
\frac{\de \sigma}{\de B_a^+ \, \de B_b^+} &= \frac{1}{2 S} \int \frac{\de^3 \vec{p}_3}{(2 \pi)^3 2 E_3}\, \int \frac{\de^3 \vec{p}_4}{(2\pi)^3 2 E_4}\,
\sum_X\!\!\!\!\!\!\!\!\!\int \, \, (2\pi)^4 \delta^{(4)}(P_a+P_b - p_3 - p_4 -p_X) \,\nonumber \\
& \quad \quad \times \frac{1}{4 d^2_R}\bigg | \bra{t(p_3) \bar{t}(p_4) X(p_X)}  \mathcal{H}_{\rm eff}(0)\ket{P_1(P_a)P_2(P_b)} \bigg|^2\,\nonumber \\
& \quad \quad \times \delta\big(B_a^+-n_a\cdot B_a(X)\big) \delta\big(B_b^+ - n_b \cdot B_b(X)\big)\, ,
\end{align}
where the factor $1/4$ is due to the average over the initial state polarisations and the factor $d_R=\{N_c,\,N_c^2-1\}$ for the average over the colours of the initial state quarks or gluons.
The hemisphere hadronic momenta $B_{a,b}^\mu(X)$ are obtained when the hemisphere momentum operators $\hat{p}^\mu_{a,b}$ act on the radiation states as
\begin{align}\label{eq:momop}
\hat{p}^\mu_{a,b} \ket{X} = B^\mu_{a,b}(X) \ket{X}\, .
\end{align}
We proceed by inserting in \eq{crosssection}  the identity
\begin{align}
1 = \int\, \de M^2 \, \de^4 q  \, \delta^{(4)}(q-p_3-p_4)\, \delta(M^2 -q^2) \, ,
\end{align}
where $q^\mu \equiv p_3^\mu + p_4^\mu$ is the four momentum of the $t \bar{t}$
pair and we denote the invariant mass and the rapidity of the \ttbar system by
$M$ and $\yttbar$ respectively.

After a few manipulations we arrive at
\begin{align}
\frac{\de \sigma}{\de B_a^+ \, \de B_b^+} = \frac{1}{16 S}  \frac{1}{(2
  \pi)^2} & \!\! \int \!\! \de M\, \de \yttbar\, \de \cos \theta \,\de
\phi_t M \beta_t \!\! \int \!\! \frac{\de^2 \vec{q}_\perp}{(2 \pi)^4}\, \sum_X\!\!\!\!\!\!\!\!\!\int\,  (2\pi)^4 \delta^{(4)}(P_a + P_b - q -p_X) \,\nonumber \\
\times  \frac{1}{4 d^2_R} \bigg | \bra{t(p_3)  \bar{t}(p_4) X(p_X)} &  \mathcal{H}_{\rm
  eff}(0)\ket{P_1(P_a)P_2(P_b)} \bigg|^2  \delta\big(B_a^+-n_a\cdot B_a(X)\big) \delta\big(B_b^+ - n_b \cdot B_b(X)\big)\, ,
\end{align}
where $\beta_t$ and $\theta$ are defined in \eq{variables} and $\phi_t$ is the azimuthal angle of the top quark ($v_3$) in the $t \bar{t}$ rest frame.
By employing the  definition of the Dirac delta function,
\begin{equation}
\delta^{(4)}(P_a+P_b - p_3 - p_4 -p_X) = \int \frac{\de^4 x}{(2 \pi)^4}   e^{i (P_a+P_b - p_3 - p_4 -p_X)\cdot x} \, ,
\end{equation}
the translation operator acting on the effective Hamiltonian 
\begin{equation}
\mathcal{H}_{\rm eff} (x) = e^{+i \hat P\cdot \hat x} \mathcal{H}_{\rm eff}(0) e^{-i \hat P\cdot \hat x } \, ,
\end{equation}
and the momentum conservation relation, we rewrite the differential cross section as
\begin{align}
&\frac{\de \sigma}{\de M\, \de \yttbar\, \de\cos \theta\,\, \de B_a^+ \, \de B_b^+} 
= \, \nn \\
& \qquad   \frac{M \beta_t}{16 S}\frac{1}{(2 \pi)^2} \, \frac{1}{4 d^2_R}\int \,\de \phi_t\, \,\int \frac{\de^2 \vec{q}_\perp}{(2 \pi)^4}\,\int \de^4 x \,  e^{+i(P_a+P_b - p_3 - p_4 -p_X)\cdot x} \,\nonumber \\
& \qquad \times \sum_X\!\!\!\!\!\!\!\!\!\int \, \,\bra{P_1(P_a) P_2(P_b)} e^{-i \hat P\cdot \hat x} \, \mathcal{H}^{\dagger}_{\rm eff}(x) \, e^{i \hat P\cdot \hat x}\ket{t(p_3) \bar{t}(p_4) X(p_X)}  \nn \\
& \qquad \quad \times  \bra{t(p_3) \bar{t}(p_4) X(p_X)}  \mathcal{H}_{\rm eff}(0)\ket{P_1(P_a)P_2(P_b)} \, \delta\big(B_a^+-n_a\cdot B_a(X)\big) \delta\big(B_b^+ - n_b \cdot B_b(X)\big)\,\nonumber \\
& \ \  =  \frac{M \beta_t}{16 S}\frac{1}{(2 \pi)^2} \, \frac{1}{4 d^2_R}\int \,\de \phi_t\,  \,\int \frac{\de^2 \vec{q}_\perp}{(2 \pi)^4}\,\nn \\
&  \qquad \times \int \de^4 x  \, \,\bra{P_1(P_a) P_2(P_b)} \, \mathcal{H}^{\dagger}_{\rm eff}(x) \, \ket{t(p_3) \bar{t}(p_4)} \delta(B_a^+-n_a\cdot \hat{p}_a ) \, \nonumber \\
& \qquad \quad \times \delta(B_b^+ - n_b \cdot \hat{p}_b)\, \bra{t(p_3) \bar{t}(p_4)}  \mathcal{H}_{\rm eff}(0)\ket{P_1(P_a)P_2(P_b)} \,,
\end{align}
where in the last line we have also expressed the delta function constraints in terms of the hemisphere momentum operators $\hat{p}^\mu_{a,b}$ which act on the radiation states as in \eq{momop}.

For the purpose of the derivation of the factorisation formula, we explicitly work out all the details for the gluon fusion channel contribution. The result for the quark anti-quark channel case can be obtained in a very similar way. After substituting the explicit expression for the second operator in \eq{operators} one obtains
\begin{align}
&\frac{\de \sigma^{gg}}{\de M\, \de \yttbar\, \de\cos \theta\,\, \de B_a^+ \, \de B_b^+} = \,  \\
&  \quad \frac{M \beta_t}{32 S} \frac{1}{(2 \pi)^4} \, \frac{1}{4 d_R^2}\, \sum_{I,I^\prime,m,m^\prime}\,  \,\int \,\de \phi_t\, \int\, \de x^+ \,  \de x^- \, e^{-i (q^- x^+ + q^+ x^-)/2}\, \nonumber \\
&   \quad \times \int \, \de^2 \vec{x}_\perp  \, \delta^{(2)}(\vec{x}_\perp) \, \int \de r^\prime\, \de t^\prime\, \de r\, \de t \,C^{gg\,\, *}_{I^\prime m^\prime}(r^\prime,t^\prime)\,\,  C^{gg}_{I m}(r,t) \,\nonumber \\
&   \quad \times \bra{P_1(P_a) P_2(P_b)}\!\!\!\!\! \sum_{\{a^\prime\},\{b^\prime\}}\!\!\!\!\!  \big(c^{gg\, *}_{I^\prime}\big)_{\{a^\prime\}} \big[O^{h\, \dagger}_{m^\prime}(x)\big]^{\rho \sigma}_{b^\prime_3 b^\prime_4} \big[O^{c\, \dagger}(x,r^\prime,t^\prime)\big]^{b^\prime_1 b^\prime_2}_{\rho\sigma} \big[O^{s\, \dagger}(x)\big]^{\{a^\prime\},\{b^\prime\}}\! \ket{t(p_3) \bar{t}(p_4)}\, \nonumber \\
&   \quad \times \delta(B_a^+-n_a\cdot \hat{p}_a ) \, \delta(B_b^+ - n_b \cdot \hat{p}_b)\, \nonumber \\
&   \quad \times \bra{t(p_3) \bar{t}(p_4)}  \sum_{\{a\},\{b\}}  \big(c^{gg}_I\big)_{\{a\}}\, \big[O^h_m(0)\big]^{\mu \nu}_{b_3 b_4} \big[O^c(0,r,t)\big]^{b_1 b_2}_{\mu\nu} \big[O^s(0)\big]^{\{a\},\{b\}}\,\ket{P_1(P_a)P_2(P_b)} \,.\nonumber
\end{align}
We  define a vector of Wilson coefficients
\begin{align}
\ket{C_m(r,t)} = \sum_I\, C_{I\, m}(r,t) \ket{c_I}\, ,
\end{align}
where the colour basis tensors in \eq{colorbasis} are understood as basis vectors $\ket{c_I}$ of an abstract colour space.
We employ the colour space formalism to simplify the notation and rewrite the formula above as
\begin{align}\label{eq:beforemultipole}
&\frac{\de \sigma^{gg}}{\de M\, \de \yttbar\, \de\cos \theta\,\, \de B_a^+ \, \de B_b^+} = \,  \\
&\qquad  \frac{M \beta_t}{32 S} \frac{1}{(2 \pi)^4} \, \frac{1}{4 d_R^2}\, \sum_{m,m^\prime}\,  \,\int \,\de \phi_t\, \int\, \de x^+ \,  \de x^- \, e^{-i (q^- x^+ + q^+ x^-)/2}\, \nonumber \\
& \quad \quad \times \int \, \de^2 \vec{x}_\perp  \, \delta^{(2)}(\vec{x}_\perp) \, \int \de r^\prime\, \de t^\prime\, \de r\, \de t  \,\nonumber \\
& \quad \quad \times \bra{C_{m^\prime}(r^\prime, t^\prime)}\bra{P_1(P_a) P_2(P_b)} \, \big[O^{h\, \dagger}_{m^\prime}(x)\big]^{\rho \sigma} \big[O^{c\,\dagger}(x,r^\prime\!,t^\prime\!)\big]_{\rho\sigma} \big[\bm{O}^{s\, \dagger}(x)\big]\, \, \ket{t(p_3) \bar{t}(p_4)}\, \nonumber \\
& \quad \quad \times \delta(B_a^+-n_a\cdot \hat{p}_a ) \, \delta(B_b^+ - n_b \cdot \hat{p}_b)\, \nonumber \\
& \quad \quad \times \bra{t(p_3) \bar{t}(p_4)}  \,\big[O^h_m(0)\big]^{\mu \nu} \big[O^c(0,r,t)\big]_{\mu\nu} \big[\bm{O}^s(0)\big] \ket{P_1(P_a)P_2(P_b)} \ket{C_m(r,t)}\,. \nonumber
\end{align}
where the soft operator $\bm{O}^s(x)$ contains soft Wilson lines $\bm{Y}_i$ with the colour generators $\bm{T}_i$ acting in abstract colour space.

After the BPS (decoupling) transformation~\cite{Bauer:2001yt} the effective Lagrangian is written in terms of independent collinear, anti-collinear, heavy quark and soft Lagrangians which do not interact with each other. For this reason and the fact that the final state heavy quarks do not contribute to the radiation in any hemisphere, we can write the hemisphere momentum operators $\hat{p}_a$ and $\hat{p}_b$ as the sum of independent operators which act in the separate collinear, anti-collinear and soft sectors as~\cite{Stewart:2009yx}
\begin{align}
\hat{p}_a = \hat{p}_{a,n_a}+ \hat{p}_{a,n_b} + \hat{p}_{a,s}\, ,\quad \quad
\hat{p}_b = \hat{p}_{b,n_a}+ \hat{p}_{b,n_b} + \hat{p}_{b,s}\, .
\end{align}

The $n_a$ collinear sector cannot contribute to the momentum in the $n_b$ hemisphere and vice versa. Therefore $\hat{p}_{a,n_b}=\hat{p}_{b,n_a}=0$ and the relations above simplify to
\begin{align}
\hat{p}_a = \hat{p}_{n_a} + \hat{p}_{a,s}\, ,\quad \quad
\hat{p}_b = \hat{p}_{n_b} + \hat{p}_{b,s}\, ,
\end{align}
where $\hat{p}_{a,n_a}=\hat{p}_{n_a}$ and $\hat{p}_{b,n_b}=\hat{p}_{n_b}$ represent the total momentum operators of the collinear and anti-collinear sectors.
We separate the soft and collinear contributions to the single hemispheres using the relations
\begin{align}
\delta(B_a^+ - n_a \cdot \hat{p}_a) &= \int \de b_a^+ \, \de k_a^+ \, \delta(B_a^+ - b_a^+ - k_a^+)\, \delta(b_a^+ - n_a \cdot \hat{p}_{n_a})\, \delta(k_a^+-n_a\cdot \hat{p}_{a,s})\, , \\
\delta(B_b^+ - n_b \cdot \hat{p}_b) &= \int \de b_b^+ \, \de k_b^+ \, \delta(B_b^+ - b_b^+ - k_b^+)\, \delta(b_b^+ - n_b \cdot \hat{p}_{n_b})\, \delta(k_b^+-n_b\cdot \hat{p}_{b,s})\, ,
\end{align}
where $b_a^\mu$, $b_b^\mu$ and $k_a^\mu$, $k_b^\mu$ are the contributions of the collinear and soft momenta from each hemisphere respectively.
We also apply the multipole expansion to  the arguments of the collinear and soft fields in $x_\mu$ so that the collinear fields are evaluated at $x^+_\mu$, the anti-collinear fields at $x^-_\mu$ and the soft fields at the origin. 
We rewrite the delta function constraints on the collinear fields in position space by introducing two integrations over the components $x_a^-$ and $x_b^-$ and
 insert the following identities
\begin{align}
1 &= e^{+i \xbm n_b\cdot \hat{p}_{n_b}/2} e^{-i \xbm n_b\cdot \hat{p}_{n_b}/2},\quad \quad 1 = e^{-i \xam n_a\cdot \hat{p}_{n_a}/2} e^{i \xam n_a\cdot \hat{p}_{n_a}/2}, \,
\end{align}
between the bra (or ket) and the collinear (anti-collinear) fields to translate the field arguments by $\xam n_a^\mu/2$ ($\xbm n^\mu_b/2$).
In addition, we use the fact that the momentum operators $n_a\cdot \hat{p}_{n_a}$ and $n_b\cdot \hat{p}_{n_b}$ acting on proton states give
\begin{equation}
n_{a}\cdot \hat{p}_{n_{a}} \ket{P_{1}(P_{a})}=0, \,
\qquad
n_{b}\cdot \hat{p}_{n_{b}} \ket{P_{2}(P_{b})}=0 \,.
\end{equation}
We obtain
\begin{align}
&\frac{\de \sigma^{gg}}{\de M\, \de \yttbar\, \de\cos \theta\,\, \de B_a^+ \, \de B_b^+} = \,  \\
& \quad \ \frac{M \beta_t}{32 S} \frac{1}{(2 \pi)^4} \, \frac{1}{4 d_R^2}\,  \int \, \de k_a^+ \, \de k_b^+\, \de b_a^+\, \de b_b^+ \,\,\delta(B_a^+ - b_a^+ - k_a^+)\, \delta(B_b^+ - b_b^+ - k_b^+)  \,\nonumber\\ & \quad \quad \times \int \,\de \phi_t\, \int\, \de x^+ \,  \de x^- \, e^{-i (q^- x^+ + q^+ x^-)/2} \, \int \de r^\prime\, \de t^\prime\, \de r\, \de t  \,\nonumber \\
& \quad \quad \times \sum_{m,m^\prime}\,\ket{C_m(r,t)}\bra{C_{m^\prime}(r^\prime, t^\prime)} \bra{0}\big[O^{h\, \dagger}_{m^\prime}(0)\big]^{\rho \sigma} \ket{t(p_3) \bar{t}(p_4)}\bra{t(p_3) \bar{t}(p_4)} \big[O^h_m(0)\big]^{\mu \nu}\ket{0}\, \nonumber \\
& \quad \quad \times \bra{P_1(P_a)}  \mathcal{A}_{c \rho \perp}(x^+ n_b/2 + t^\prime n_b)\,\delta(b_a^+ - n_a\cdot \hat{p}_{n_a})\, \mathcal{A}_{c \mu \perp}(t n_b)\ket{P_1(P_a)}  \, \nonumber \\
& \quad \quad \times   \bra{P_2(P_b)} \mathcal{A}_{\bar{c} \sigma \perp}(x^- n_a/2 + r^\prime n_a)\, \delta(b_b^+ - n_b\cdot \hat{p}_{n_b}) \,\mathcal{A}_{\bar{c} \nu \perp}(r n_a) \ket{P_2(P_b)} \, \nonumber \\
& \quad \quad \times \bra{0}\bar{\mathbf{T}}\big[\bm{O}^{s\, \dagger}(0)\big]\, \delta(k_a^+-n_a\cdot \hat{p}_{a,s})\,\delta(k_b^+-n_b\cdot \hat{p}_{b,s}) \, \mathbf{T} \big[\bm{O}^s(0)\big]  \ket{0}\,\nonumber \\
& = \frac{M \beta_t}{32 S} \frac{1}{(2 \pi)^4} \, \frac{1}{4 d_R^2}\,  \int \, \de k_a^+ \, \de k_b^+\, \de b_a^+\, \de b_b^+ \,\,\delta(B_a^+ - b_a^+ - k_a^+)\, \delta(B_b^+ - b_b^+ - k_b^+)  \,\nonumber\\ & \quad \quad \times \int \,\de \phi_t\, \int\, \de x^+ \,  \de x^- \, e^{-i (q^- x^+ + q^+ x^-)/2} \, \int \de r^\prime\, \de t^\prime\, \de r\, \de t  \,\nonumber \\
& \quad \quad \times \sum_{m,m^\prime}\,\ket{C_m(r,t)}\bra{C_{m^\prime}(r^\prime, t^\prime)} \bra{0}\big[O^{h\, \dagger}_{m^\prime}(0)\big]^{\rho \sigma} \ket{t(p_3) \bar{t}(p_4)}\bra{t(p_3) \bar{t}(p_4)} \big[O^h_m(0)\big]^{\mu \nu}\ket{0}\, \nonumber \\
& \quad \quad \times \bigg(\int \frac{\de x^-_{a}}{4 \pi} e^{+i x^-_{a} b_a^+/2}\bra{P_1(P_a)}  \mathcal{A}_{c \rho \perp}(x^+ n_b/2 + \xam n_a/2+ t^\prime n_b)\,\mathcal{A}_{c \mu \perp}(t n_b)\ket{P_1(P_a)}  \bigg) \, \nonumber \\
& \quad \quad \times \bigg(\int \frac{\de x^-_{b}}{4 \pi} e^{+i x^-_{b} b_b^+/2} \bra{P_2(P_b)} \mathcal{A}_{\bar{c} \sigma \perp}(x^- n_a/2 + \xbm n_b/2 + r^\prime n_a)\,\mathcal{A}_{\bar{c} \nu \perp}(r n_a) \ket{P_2(P_b)} \bigg) \, \nonumber \\
& \quad \quad \times \bigg(\sum_{X_s}\!\!\!\!\!\!\!\!\!\int\, \bra{0}\bar{\mathbf{T}}\big[\bm{O}^{s\, \dagger}(0)\big]\ket{X_s}\bra{X_s}\mathbf{T} \big[\bm{O}^s(0)\big]  \ket{0} \delta(k_a^+ - k_a^+(X_s)) \, \delta(k_b^+  - k_b^+(X_s))\bigg)\,. \nonumber
\end{align}
In the last line of the equation above we have also introduced the eigenvalues $k_{i}^+(X_s) $ of the projection of the momentum operator on a soft state $\ket{X_s}$ via $ n_i \cdot \hat{p}_{i,s} \ket{X_s} = k_{i}^+(X_s) \ket{X_s}$.
At this point we introduce an operatorial definition for the gluon beam function $B_{g/P_1}$ as follows
\begin{align}\label{eq:beamdef}
\bra{P_1(P_a)}  &\mathcal{A}^a_{c \rho \perp}(x^+ \bar{n}_a/2 + \xam n_a/2 + t^\prime \bar{n}_a)  \mathcal{A}^b_{c \mu \perp}(t \bar{n}_a)\ket{P_1(P_a)} = \, \nonumber \\
& (-g_{\rho\mu \perp}) \,\delta^{ab} \, (\bar{n}_a \cdot P_a)\,\int^1_0 \de z_a \, e^{i z_a [x^+ \bar{n}_a/2+(t^\prime-t)\bar{n}_a]\cdot P_a}\, \int \de b^{+\,\prime}_a e^{-i (x^-_{a} b^{+\, \prime}_a)/2} \, \nonumber \\ & \quad \quad\times B_{g/P_1}\big(z_a (\bar{n}_a\cdot P_a)\, b^{+\, \prime}_a,z_a,\mu\big) \,.
\end{align}
Notice that the Fourier transform which appears in the equation above misses a factor $1/(4 \pi)$ compared to the standard Fourier transform conventions in Ref.~\cite{Becher:2014oda}. We do this in order to normalise the tree level beam functions in the same way as in Ref.~\cite{Stewart:2009yx}, namely $B_i^{(0)}(t,z,\mu) = \delta(t)\,  f_i(z,\mu)$ where the $f_i$ are the standard parton distribution functions.
We also define the soft functions
\begin{align}
  \label{eq:softoper}
\mathbf{S}_{gg}&(k_a^+,k_b^+,\betat,\theta,\mu) = \, \\
&\frac{1}{d_R} \,\bigg(\sum_{X_s}\!\!\!\!\!\!\!\!\!\int\, \bra{0}\bar{\mathbf{T}}\big[\bm{O}^{s\, \dagger}(0)\big]\ket{X_s}\bra{X_s}\mathbf{T} \big[\bm{O}^s(0)\big]  \ket{0} \delta(k_a^+ - k_a^+(X_s)) \, \delta(k_b^+  - k_b^+(X_s))\bigg) \, ,\nonumber
\end{align}
and hard functions
\begin{align}
\mathbf{H}^{\mu\nu\rho \sigma}_{gg}(M,\betat,v_3,\mu) =& \frac{1}{(4 \pi)^2}\frac{1}{4 d_R} \sum_{m,m^\prime} \ket{\tilde{C}_m}\bra{\tilde{C}_{m^\prime}}\, \, \nonumber \\
& \times \bra{0}\big[O^{h\, \dagger}_{m^\prime}(0)\big]^{\rho \sigma} \ket{t(p_3) \bar{t}(p_4)}\bra{t(p_3) \bar{t}(p_4)} \big[O^h_m(0)\big]^{\mu \nu}\ket{0}\, ,
\end{align}
where
\begin{align}
\ket{\tilde{C}_m} \equiv \ket{\tilde{C}_m(M,\betat,\theta,\mu)} = \int \de r \, \de t\,  e^{-i\, z_a t\, n_b\cdot P_a} \, e^{-i\, z_b r\, n_a\cdot P_b}\,  \ket{C_m(r,t)}\, .
\end{align}

Since the two beam functions are proportional to $g^{\rho \mu}_\perp$ and $g^{\sigma \nu}_\perp$ and are contracted with the open Lorentz indices of the hard functions,  we can define a scalar hard function as
\begin{align}
\mathbf{H}_{gg}(M,\betat,\theta,\mu) = (-g_{\rho \mu\, \perp})\, (-g_{\sigma \nu\, \perp})\, \mathbf{H}^{\mu\nu\rho \sigma}_{gg}(M,\betat,v_3,\mu) \, .
\end{align}
Furthermore, we integrate over $\phi_t$  because the integrand is independent of this.
After these substitutions and integration over $\xam\,,\xbm$ the result for the differential cross section in the gluon channel reads
\begin{align}
&\frac{\de \sigma^{gg}}{\de M\, \de \yttbar\, \de\cos \theta\,\, \de B_a^+ \, \de B_b^+} = \, \nonumber \\
& \quad \frac{\beta_t M}{32 S} \frac{1}{(2 \pi)^3}  \,  \int \, \de k_a^+ \, \de k_b^+ \, \int \, \de b_a^+\, \de b_b^+ \,\,\delta(B_a^+ - b_a^+ - k_a^+)\, \delta(B_b^+ - b_b^+ - k_b^+)  \,\nonumber\\
& \quad \times S \, \int \de z_a\, \de z_b \int\, \de x^+ \,  \de x^- \, e^{-i (q^- x^+ + q^+ x^-)/2}\,\, e^{i z_a x^+ n_b\cdot P_a/2} \, e^{i z_b x^- n_a\cdot P_b/2}\, \nonumber \\
& \quad \times \int \de r^\prime\, \de t^\prime\, \de r\, \de t \, \,e^{i z_a (t^\prime-t) n_b\cdot P_a} \, e^{i z_b (r -r^\prime) n_a\cdot P_b}\,  \,\nonumber \\
& \quad \times \big[(-g_{\rho \mu \, \perp}) \, B_{g/P_1}\big(z_a (n_b\cdot P_a) \, b_a^+,z_a,\mu\big) \, (-g_{\sigma \nu \, \perp})  \, B_{g/P_2}\big(z_b (n_a\cdot P_b) \, b^+_b,z_b,\mu\big) \big] \, \nn \\
& \quad \times  \frac{1}{4 d_R} \,\mathrm{Tr}\Bigg\{\!\! \sum_{m,m^\prime} \!\! \ket{C_m(r,t)}\bra{C_{m^\prime}(r^\prime, t^\prime)} \bra{0}\big[O^{h\, \dagger}_{m^\prime}(0)\big]^{\rho \sigma} \ket{t(p_3) \bar{t}(p_4)}\bra{t(p_3) \bar{t}(p_4)} \big[O^h_m(0)\big]^{\mu \nu}\ket{0}\, \nonumber \\
& \qquad \qquad  \qquad  \qquad \mathbf{S}_{gg}(k^+_a,k^+_b,\betat, \theta,\mu) \Bigg\}\,\nonumber \\
& =  \frac{\pi \beta_t M}{S}   \,  \int \, \de k_a^+ \, \de k_b^+\, \int \, \de b_a^+\, \de b_b^+ \,\,\delta(B_a^+ - b_a^+ - k_a^+)\, \delta(B_b^+ - b_b^+ - k_b^+)  \,\nonumber\\
& \quad \quad  \times \int \de z_a\, \de z_b \, \delta\bigg(z_a -\frac{q^-}{n_b\cdot P_a}\bigg) \, \delta\bigg(z_b - \frac{q^+}{n_a \cdot P_b}\bigg)\, \nonumber \\
&  \quad \quad  \times \big[ B_{g/P_1}\big(z_a (n_b\cdot P_a) \, b_a^+,z_a,\mu\big) \, B_{g/P_2}\big(z_b (n_a\cdot P_b) \, b^+_b,z_b,\mu\big) \big] \,  \nonumber \\
& \quad \quad \times \mathrm{Tr}\big\{ \mathbf{H}_{gg}(M,\betat,\theta,\mu)\, \mathbf{S}_{gg}(k_a^+,k_b^+,\betat, \theta,\mu)\big\} \,\nonumber \\
&= \frac{\pi \beta_t M}{S} \, \int \, \de k^+_a \, \de k^+_b \,  \mathrm{Tr}\big\{ \mathbf{H}_{gg}(M,\betat,\theta,\mu)\, \mathbf{S}_{gg}(k_a^+,k_b^+,\betat, \theta,\mu)\big\} \nn \\
& \qquad \times \big[B_{g/P_1}\big(q^-(B^+_a - k^+_a),q^-/n_b\cdot P_a,\mu\big) \, B_{g/P_2}\big(q^+ (B^+_b - k^+_b), q^+/n_a\cdot P_b,\mu\big)\big] \,,
\end{align}
where $q^\mp = M \, e^{\pm \yttbar}$ and the Kronecker deltas with colour indices arising from the beam functions produce a colour trace over the product of the hard and soft functions.

We now wish to define a variable to constrain both hemispheres simultaneously. When the rapidity of the \ttbar system is zero, the hemispheres are equal in size and a good choice is
\begin{align}
\label{eq:bhatdef}
    \widehat{B}=\frac{B_a^+ +B_b^+}{M} \, .
\end{align}
Integrating over $B_a^+$ and $B_b^+$ we obtain
\begin{align}
&\frac{\de \sigma^{gg}}{\de M\, \de \yttbar\, \de\cos \theta\,\, \de \widehat{B}} = \frac{\pi \beta_t M}{S} \, \int \, \de k^+_a \, \de k^+_b \, \de B_a^+\,\de B_b^+ \, \delta(M\widehat{B}-B_a^+-B_b^+)\nonumber \\
&\qquad \quad\times\big[B_{g/P_1}\big(q^-(B^+_a - k^+_a),q^-/n_b\cdot P_a,\mu\big) \, B_{g/P_2}\big(q^+ (B^+_b - k^+_b), q^+/n_a\cdot P_b,\mu\big)\big] \, \nonumber \\
&\qquad \quad\times\, \mathrm{Tr}\big\{ \mathbf{H}_{gg}(M,\betat,\theta,\mu)\, \mathbf{S}_{gg}(k_a^+,k_b^+,\betat, \theta,\mu)\big\} \, \nonumber\\
&\quad =\frac{\pi \beta_t }{MS} \, \int \, \de t_a\,\de t_b \, \big[B_{g/P_1}\big(t_a, z_a,\mu\big) \, B_{g/P_2}\big(t_b, z_b,\mu\big)\big]\nonumber \\
&\qquad \quad\times \, \mathrm{Tr}\Bigg\{ \mathbf{H}_{gg}(M,\betat,\theta,\mu)\, \mathbf{S}_{B,gg}\left(M\widehat{B}-\frac{t_a}{q^-}-\frac{t_b}{q^+},\betat, \theta,\mu\right)\Bigg \}\,,
\label{eq:penultimateform}
\end{align}
where we have used that $B_{a,b}^+=k_{a,b}^++b_{a,b}^+$, $t_{a}=q^{-} b_{a}^+$ and $t_{b}= q^+ b_b^+$, and have defined the soft function as
\begin{align}
\mathbf{S}_{B,gg}(\tausq,\betat, \theta,\mu) = \int \de k_a^+\, \de k_b^+\,\mathbf{S}_{gg}(k_a^+,k_b^+,\betat, \theta,\mu)\,\delta(\tausq - k_a^+-k_b^+)\,.
\end{align}
Boosting the \ttbar system to a generic rapidity $\yttbar$ will, however, alter not only the distribution of the QCD radiation but also change the definition of the hemispheres. Defining rapidity-dependent hemispheres $a$ and $b$ for $y>\yttbar$ and $y<\yttbar$, we introduce the boost-invariant combination
\begin{align}
\tau_B=\frac{q^- B_a^+(\yttbar)+q^+B_b^+(\yttbar)}{M^2}\,.
\label{eq:taudef}
\end{align}
In the partonic centre-of-mass frame, the definition of $\tau_B$ in \eq{taudef} coincides with the definition of $\widehat{B}$ in \eq{bhatdef} and one can replace $\widehat{B}$ in \eq{penultimateform} with $\tau_B$. Boosting to a generic frame at rapidity $\yttbar \neq 0$, the (invariant) beam functions are unaffected: the soft function, on the other hand, is boost-invariant up to the hemisphere definition which defines its arguments $k^+_{a,b}$. It can easily be shown~\cite{Stewart:2009yx} that the simultaneous transformation of the soft function and its arguments results in the soft function for $\tau_B$ being the same as for $\widehat{B}$ -- we can therefore write 
\begin{align}
\frac{\de \sigma^{gg}}{\de M\, \de \yttbar\, \de\cos \theta\,\, \de \tau_B} &= \frac{\pi \beta_t }{MS} \, \int \, \de t_a\,\de t_b \, \big[B_{g/P_1}\big(t_a, z_a,\mu\big) \, B_{g/P_2}\big(t_b, z_b,\mu\big)\big] \\
&\qquad\times \, \mathrm{Tr}\Bigg\{ \mathbf{H}_{gg}(M,\betat,\theta,\mu)\, \mathbf{S}_{B,gg}\left(M\tau_B-\frac{t_a+t_b}{M},\betat,\theta,\mu\right)\Bigg \}\,.\nn
\label{eq:finalform}
\end{align}

The derivation of the factorisation formula for the quark channel follows similarly to the gluon case. The main difference involves the operatorial definition of the quark beam function $ B_{q/P_1}$ which is given by
\begin{align}
\bra{P_1(P_a)} \bar{\chi}_{c}&(x^+ \bar{n}_a/2 + \xam n_a/2 + t^\prime \bar{n}_a)   \, \frac{\slashed{n}_b}{2} \chi_{c}( t \bar{n}_a) \ket{P_1(P_a)} = \, \nonumber \\
& \quad (\bar{n}_a\cdot P_a)^2\,  \int^1_0 \, \de z_a \, z_a\, e^{i z_a [x^+ \bar{n}_a/2 + (t^\prime-t) \bar{n}_a ]\cdot P_a}\, \int \, \de b_a^+\, e^{-i (x^-_{a} b_a^+)/2}\,\nonumber \\
&\quad\quad \quad \times  B_{q/P_1}\big(z_a (\bar{n}_a\cdot P_a) b_a^+,z_a,\mu\big)\, .
\end{align}

\section{One-loop soft integrals}
\label{app:softints}
In this appendix we collect the integrals that need to be evaluated for the one-loop soft functions. We have analytically calculated the pole structure of all integrals. The terms at finite order in the regulator $\epsilon$, however, resisted our attempts at an analytic calculation, leaving several finite one-fold integrals to be performed numerically.~\footnote{Analytic results for the finite terms in $\epsilon$ of similar integrals have been presented in Refs.~\cite{Somogyi:2011ir,Lyubovitskij:2021ges}. However, in our case the differing phase space constraints mean that we are unable to make use of these results; specifically, since we integrate over each hemisphere separately, we introduce a step function in the cosine of the angle $\chi$. This is absent in the aforementioned works.}

We parameterise the momenta of the soft parton and of the top and anti-top quarks as
\begin{align}
k & = k_0\, (1,\ldots,\sin \chi \sin \phi,\sin \chi \cos \phi, \cos \chi),\,\\
v_3 & = \frac{1}{\sqrt{1-\beta^2_t}}\, (1, \ldots,0, \beta_t \sin \theta, \beta_t \cos \theta) ,\, \\
v_4 & =  \frac{1}{\sqrt{1-\beta^2_t}}\, (1, \ldots, 0, - \beta_t \sin \theta, - \beta_t \cos \theta)   , \,
\end{align}
respectively. We have $\beta_t = \sqrt{1-4 m^2_t/M^2}$ and $M$ is the invariant mass of the $t \bar{t}$ pair.
The integration measure is
\begin{align}
\de^d k \, \delta(k^2)\Theta(k^0) =  \Omega_{d-4} \frac{k^{1-2 \epsilon}_0}{2}\, \de k_0 \sin^{1-2 \epsilon} \chi \,\de \chi \,  \sin^{-2 \epsilon}\phi\,  \de \phi\, ,
\end{align}
and the volume of the $d-4 = - 2\epsilon$ space is given by
\begin{equation}
    \Omega_{d-4}=\frac{2\pi^{1/2-\epsilon}}{\Gamma(\frac{1}{2}-\epsilon)}=\frac{2^{1-2\epsilon}\pi^{-\epsilon}\Gamma(1-\epsilon)}{\Gamma(1-2\epsilon)} \, .
\end{equation}
We also define the variable $z=\cos\chi$. In the following we collect the relevant integrals, separated  according to the configuration of massless and massive legs and their directions with respect to the hemispheres.

\subsection{Heavy-heavy integrals, same leg}
\begin{align}
    I_{v_3v_3}&=\int \de^dk \delta(k^2)\Theta(k_0) \frac{v_3\cdot v_3}{(v_3\cdot k)^2} \delta(\tausq-k\cdot n_a)\Theta(k\cdot n_b - k\cdot n_a) \nn \\
    &=\frac{1}{2}\dvol \int \de k_0 \de\chi \de\phi\, k_0^{1-2\epsilon}\sin^{1-2\epsilon}\chi \sin^{-2\epsilon}\phi \frac{1-\beta_t^2}{k_0^2}\nn\\&\qquad \qquad \times (1-\beta_t\sin\chi\cos\phi\sin\theta-\beta_t\cos\chi\cos\theta)^{-2}  \delta(\tausq-k_0(1-\cos\chi))\Theta(\cos\chi)\nn \\
    &=\frac{1}{2}\dvol\int_0^\pi \de\phi \int_0^1 \de z \,(\tausq)^{-1-2\epsilon} (1-\beta_t^2) \left(\frac{1-z}{1+z}\right)^\epsilon \sin^{-2\epsilon}\phi \nn\\&\qquad\qquad\times(1-\beta_t\sqrt{1-z^2}\cos\phi\sin\theta - \beta_t z\cos\theta)^{-2}\nn\\&=\frac{\pi}{2}\dvol(\tausq)^{-1-2\epsilon} \left(1+\frac{\beta_t\cos\theta}{\sqrt{1-\beta_t^2\sin^2\theta}}+\mathcal{O}(\epsilon)\right)
\end{align}

\begin{align}
    I_{v_4v_4}&=\int \de^d k \delta(k^2)\Theta(k_0) \frac{v_4\cdot v_4}{(v_4\cdot k)^2} \delta(\tausq-k\cdot n_a)\Theta(k\cdot n_b - k\cdot n_a) \nn \\
    &=\frac{1}{2}\dvol\int \de k_0 \de\chi \de\phi\, k_0^{1-2\epsilon}\sin^{1-2\epsilon}\chi \sin^{-2\epsilon}\phi \frac{1-\beta_t^2}{k_0^2} \nn\\&\qquad \qquad \times (1+\beta_t\sin\chi\cos\phi\sin\theta+\beta_t\cos\chi\cos\theta)^{-2} \delta(\tausq-k_0(1-\cos\chi))\Theta(\cos\chi)\nn \\
    &=\frac{1}{2}\dvol\int_0^\pi \de\phi \int_0^1 \de z \,(\tausq)^{-1-2\epsilon} (1-\beta_t^2) \left(\frac{1-z}{1+z}\right)^\epsilon \sin^{-2\epsilon}\phi \nn\\&\qquad\qquad\times(1+\beta_t\sqrt{1-z^2}\cos\phi\sin\theta + \beta_t z\cos\theta)^{-2}\nn\\&=\frac{\pi}{2}\dvol(\tausq)^{-1-2\epsilon} \left(1-\frac{\beta_t\cos\theta}{\sqrt{1-\beta_t^2\sin^2\theta}}+\mathcal{O}(\epsilon)\right)
\end{align}

\subsection{Heavy-heavy integrals, opposite leg}
\begin{align}
    I_{v_3v_4}&=\int \de^d k \delta(k^2)\Theta(k_0) \frac{v_3\cdot v_4}{(v_3\cdot k)(v_4\cdot k)} \delta(\tausq-k\cdot n_a)\Theta(k\cdot n_b - k\cdot n_a)\nn \\
    &=\frac{1}{2}\dvol\int \de k_0 \de\chi \de\phi\, k_0^{1-2\epsilon}\sin^{1-2\epsilon}\chi \sin^{-2\epsilon}\phi \frac{1+\beta_t^2}{k_0^2} \nn\\&\qquad \qquad \times [1-\beta_t^2(\sin\chi\cos\phi\sin\theta+\beta_t\cos\chi\cos\theta)^2]^{-1} \delta(\tausq-k_0(1-\cos\chi))\Theta(\cos\chi)\nn \\
    &=\frac{1}{2}\dvol\int_0^\pi \de\phi \int_0^1 \de z \,(\tausq)^{-1-2\epsilon} (1+\beta_t^2) \left(\frac{1-z}{1+z}\right)^\epsilon \sin^{-2\epsilon}\phi \nn\\&\qquad\qquad\times[1-\beta_t^2(\sqrt{1-z^2}\cos\phi\sin\theta + \beta_t z\cos\theta)^2]^{-1}\nn\\&=\frac{\pi}{2}\dvol(\tausq)^{-1-2\epsilon}\frac{1+\beta_t^2}{2\beta_t} \bigg[\ln\left(\frac{1+\beta_t}{1-\beta_t}\right) + \mathcal{O}(\epsilon)\bigg]
\end{align}

\begin{equation}
    I_{v_4 v_3}=I_{v_3v_4}
\end{equation}

\subsection{Heavy-light integrals, same hemisphere}
\begin{align}
    I_{n_av_3}&=\int \de^d k \delta(k^2)\Theta(k_0) \frac{n_a\cdot v_3}{(n_a\cdot k)(v_3\cdot k)} \delta(\tausq-k\cdot n_a)\Theta(k\cdot n_b - k\cdot n_a)\nn \\
    &=\frac{1}{2}\dvol\int \de k_0 \de\chi \de\phi\, k_0^{1-2\epsilon}\sin^{1-2\epsilon}\chi \sin^{-2\epsilon}\phi \frac{1-\beta_t\cos\theta}{k_0^2(1-\cos\chi)} \nn\\&\qquad \qquad \times (1-\beta_t\sin\chi\cos\phi\sin\theta-\beta_t\cos\chi\cos\theta)^{-1} \delta(\tausq-k_0(1-\cos\chi))\Theta(\cos\chi)\nn \\
    &=\frac{1}{2}\dvol\int_0^\pi \de\phi \int_0^1 \de z \,(\tausq)^{-1-2\epsilon} (1-\beta_t\cos\theta) (1-z)^{-1+\epsilon}(1+z)^{-\epsilon} \sin^{-2\epsilon}\phi \nn\\&\qquad\qquad\times(1-\beta_t\sqrt{1-z^2}\cos\phi\sin\theta - \beta_t z\cos\theta)^{-1}\nn\\&=\frac{\pi}{2}\dvol(\tausq)^{-1-2\epsilon}\bigg\{\frac{1}{\epsilon}+3\ln(2)+2\ln(1-\beta_t\cos\theta)\\ &\qquad\qquad -\ln\Big[(1-\beta_t\cos\theta)\left(2+\sqrt{4-2\beta_t^2+2\beta_t^2\cos2\theta}\right)-2\beta_t^2\sin^2\theta\Big]+\mathcal{O}(\epsilon)\bigg\}\nn
\end{align}
\begin{align}
    I_{n_b v_4}&=\int \de^d k \delta(k^2)\Theta(k_0) \frac{n_b \cdot v_4}{(n_b \cdot k)(v_4\cdot k)} \delta(\tausq-k\cdot n_a)\Theta(k\cdot n_b - k\cdot n_a)\nn \\
    &=\frac{1}{2}\dvol\int \de k_0 \de\chi \de\phi\, k_0^{1-2\epsilon}\sin^{1-2\epsilon}\chi \sin^{-2\epsilon}\phi \frac{1-\beta_t\cos\theta}{k_0^2(1+\cos\chi)} \nn\\&\qquad \qquad \times (1+\beta_t\sin\chi\cos\phi\sin\theta+\beta_t\cos\chi\cos\theta)^{-1} \delta(\tausq-k_0(1-\cos\chi))\Theta(\cos\chi)\nn \\
    &=\frac{1}{2}\dvol\int_0^\pi \de\phi \int_0^1 \de z \,(\tausq)^{-1-2\epsilon} (1-\beta_t\cos\theta) (1+z)^{-1-\epsilon}(1-z)^{\epsilon} \sin^{-2\epsilon}\phi \nn\\&\qquad\qquad\times(1+\beta_t\sqrt{1-z^2}\cos\phi\sin\theta+\beta_t z\cos\theta)^{-1}\nn\\&=\frac{\pi}{2}\dvol(\tausq)^{-1-2\epsilon}\bigg\{-\ln(2)-\ln(1-\beta_t^2)\nn\\&\qquad+\ln\Big[(1-\beta_t\cos\theta)\left(2+\sqrt{4-2\beta_t^2+2\beta_t^2\cos2\theta}\right)-2\beta_t^2\sin^2\theta\Big] +\mathcal{O}(\epsilon)  \bigg\}
\end{align}

\subsection{Heavy-light integrals, opposite hemisphere}
\begin{align}
    I_{n_av_4}&=\int \de^d k \delta(k^2)\Theta(k_0) \frac{n_a\cdot v_4}{(n_a\cdot k)(v_4\cdot k)} \delta(\tausq-k\cdot n_a)\Theta(k\cdot n_b - k\cdot n_a)\nn \\
    &=\frac{1}{2}\dvol\int \de k_0 \de\chi \de\phi\, k_0^{1-2\epsilon}\sin^{1-2\epsilon}\chi \sin^{-2\epsilon}\phi \frac{1+\beta_t\cos\theta}{k_0^2(1-\cos\chi)}\nn\\&\qquad \qquad \times (1+\beta_t\sin\chi\cos\phi\sin\theta+\beta_t\cos\chi\cos\theta)^{-1}  \delta(\tausq-k_0(1-\cos\chi))\Theta(\cos\chi) \nn \\
    &=\frac{1}{2}\dvol\int_0^\pi \de\phi \int_0^1 \de z \,(\tausq)^{-1-2\epsilon} (1+\beta_t\cos\theta) (1-z)^{-1+\epsilon}(1+z)^{-\epsilon} \sin^{-2\epsilon}\phi \nn\\&\qquad\qquad\times(1+\beta_t\sqrt{1-z^2}\cos\phi\sin\theta + \beta_t z\cos\theta)^{-1}\nn\\&=\frac{\pi}{2}\dvol(\tausq)^{-1-2\epsilon}\bigg\{\frac{1}{\epsilon}+3\ln(2)+2\ln(1+\beta_t\cos\theta)\nn\\ &\qquad-\ln\Big[(1+\beta_t\cos\theta)\left(2+\sqrt{4-2\beta_t^2+2\beta_t^2\cos2\theta}\right)-2\beta_t^2\sin^2\theta\Big]+\mathcal{O}(\epsilon)\bigg\}
\end{align}

\begin{align}
    I_{n_b  v_3}&=\int \de^d k \delta(k^2)\Theta(k_0) \frac{n_b \cdot v_3}{(n_b \cdot k)(v_3\cdot k)} \delta(\tausq-k\cdot n_a)\Theta(k\cdot n_b - k\cdot n_a) \nn \\
    &=\frac{1}{2}\dvol\int \de k_0 \de\chi \de\phi\, k_0^{1-2\epsilon}\sin^{1-2\epsilon}\chi \sin^{-2\epsilon}\phi \frac{1+\beta_t\cos\theta}{k_0^2(1+\cos\chi)} \nn\\&\qquad \qquad \times (1-\beta_t\sin\chi\cos\phi\sin\theta-\beta_t\cos\chi\cos\theta)^{-1} \delta(\tausq-k_0(1-\cos\chi))\Theta(\cos\chi) \nn \\
    &=\frac{1}{2}\dvol\int_0^\pi \de\phi \int_0^1 \de z \,(\tausq)^{-1-2\epsilon} (1+\beta_t\cos\theta) (1+z)^{-1-\epsilon}(1-z)^{\epsilon} \sin^{-2\epsilon}\phi \nn\\&\qquad\qquad\times(1-\beta_t\sqrt{1-z^2}\cos\phi\sin\theta-\beta_t z\cos\theta)^{-1}\nn\\&=\frac{\pi}{2}\dvol(\tausq)^{-1-2\epsilon}\bigg\{-\ln(2)-\ln(1-\beta_t^2)\nn\\&\qquad+\ln\Big[(1+\beta_t\cos\theta)\left(2+\sqrt{4-2\beta_t^2+2\beta_t^2\cos2\theta}\right)-2\beta_t^2\sin^2\theta\Big] +\mathcal{O}(\epsilon)  \bigg\}
\end{align}

\newpage
\bibliographystyle{JHEP}
\bibliography{geneva}

\end{document}